\documentclass[aps,prd,twocolumn,superscriptaddress,floatfix,nofootinbib]{revtex4}
\usepackage{longtable}
\usepackage{bm}
\usepackage{color}
\usepackage{graphicx}
\newcommand{\be}{\begin{equation}}
\newcommand{\ee}{\end{equation}}
\newcommand{\een}{\end{subequations}}
\newcommand{\ben}{\begin{subequations}}
\newcommand{\beq}{\begin{eqalignno}}
\newcommand{\eeq}{\end{eqalignno}}

\newcommand{\lsim}{\mathrel{\mathop{\kern 0pt \rlap
      {\raise.2ex\hbox{$<$}}}\lower.9ex\hbox{\kern-.190em $ \sim$}}}
\newcommand{\gsim}{\mathrel{\mathop{\kern 0pt
      \rlap{\raise.2ex\hbox{$>$}}}\lower.9ex\hbox{\kern-.190em $\sim$}}}

\begin{document}	

\title{Effective scalar four-fermion interaction for Ge--phobic
  exothermic dark matter and the CDMS-II Silicon excess}
\author{S. Scopel}
\affiliation{Department of Physics, Sogang University, 
Seoul, Korea, 121-742}
\author{Jong-Hyun Yoon}
\affiliation{Department of Physics, Sogang University, 
Seoul, Korea, 121-742}

\date{\today}

\begin{abstract}
We discuss within the framework of effective four--fermion
  scalar interaction the phenomenology of a Weakly Interacting Massive
  Particle (WIMP) Dirac Dark Matter candidate which is exothermic
  (i.e. is metastable and interacts with nuclear targets
  down--scattering to a lower--mass state) and $Ge$--phobic
  (i.e. whose couplings to quarks violate isospin symmetry leading to
  a suppression of its cross section off Germanium targets).  We
  discuss the specific example of the CDMS--II Silicon three-candidate
  effect showing that a region of the parameter space of the model
  exists where WIMP scatterings can explain the excess in compliance
  with other experimental constraints, while at the same time the Dark
  Matter particle can have a thermal relic density compatible with
  observation. In this scenario the metastable state $\chi$ and the
  lowest--mass one $\chi^{\prime}$ have approximately the same density
  in the present Universe and in our Galaxy, but direct detection
  experiments are only sensitive to the down--scatters of $\chi$ to
  $\chi^{\prime}$. We include a discussion of the recently calculated
  Next--to--Leading Order corrections to Dark Matter--nucleus
  scattering, showing that their impact on the phenomenology is
  typically small, but can become sizable in the same parameter space
  where the thermal relic density is compatible to observation.
\end{abstract}


\maketitle



\maketitle

\section{Introduction}
\label{sec:introduction}

Weakly Interacting Massive Particles (WIMPs) are the most popular
candidates to provide the Dark Matter (DM) that is known to make up 27
\% of the total mass density of the Universe\cite{planck} and believed
to dominate the dark halo of our Galaxy.  Many experiments are
presently trying to search for the tiny recoil energy deposited by the
elastic scattering of WIMPs off the nuclei of low--background
detectors. Some of them
(DAMA\cite{dama},CoGeNT\cite{cogent_modulation}\footnote{For a
  critical independent assessment of the CoGeNT spectral excess,
  claiming a much less significant residual effect than the official
  analysis, see \cite{cogent_davis}}) claim to observe a possible
yearly modulation effect which is expected in the signal due to the
Earth's rotation around the Sun, while
others(CoGeNT\cite{cogent_spectral}, CDMS-$Si$ \cite{cdms_si}, CRESST
\cite{cresst}) report a possibly WIMP--induced excess in their
time--averaged event spectra in tension with background estimates.
However, the excitation triggered by the latter results has been
considerably quenched by the outcome of many other experiments which
do not report any discrepancy with the estimated background:
(LUX\cite{lux}, XENON100\cite{xenon100},
XENON10\cite{xenon10},KIMS\cite{kims,kims_modulation},
CDMS-$Ge$\cite{cdms_ge}, CDMSlite \cite{cdms_lite},
SuperCDMS\cite{super_cdms}).

A peculiar feature of the experiments listed above is that, among those
publishing exclusions, the nature of the used target nuclei and the
range of the observed recoil energies never exactly overlap with those of
the experiments claiming detection: as a consequence,
the comparison among the former and the latter, and so the claim of a
discrepancy between them always involves some degree of model--dependence,
which rests in two main ingredients: the velocity distribution $f(\vec{v})$
of the incoming WIMPS and the scaling law among different targets
of the WIMP--nucleus cross section.
Traditionally, these two ingredients have been fixed to specific choices,
namely a Maxwellian velocity distribution whose r.m.s. velocity is related to
the galactic rotational velocity by hydrostatic equilibrium
(the so--called Isothermal Sphere Model) and a fermionic DM candidate with a scalar
effective coupling to quarks suppressed by the scale $M_{*}$:

\begin{equation}
 {\cal L}\ni \frac{1}{M_{*}^2}\bar{\chi}\chi \bar{q} q,
\label{eq:eff_lagrangian}
\end{equation}

\noindent inducing the same
scattering amplitude $f_p$ on protons and $f_n$ on neutrons and, as a consequence,
a total DM--nucleus cross section scaling with the the square of the atomic mass number $A$, i.e.:

\begin{equation}
\tilde{A}=Z+(A-Z)\frac{f_n}{f_p}=A,
\label{eq:scaling_law}
\end{equation}

\noindent with $Z$ the nuclear atomic number. If these assumptions are made,
indeed the experimental results listed above are in sometimes strong tension with each
other, at least when they are taken at face value and the many possible sources of
systematic errors\cite{collar_liquid} (connected to quenching factors, atomic form
factors, background cuts efficiencies, etc.) are not factored in.

In light of the situation summarized above several new directions have
been explored in the recent past both to remove as much as possible
the dependence on specific theoretical assumptions from the analysis
of DM direct detection data and to extend its scope to a wider class
of models.  Starting from \cite{factorization}, at least for
experiments not involving annual modulation\footnote{The extension of
  the halo--independent approach to the annually modulated part of the
  expected rate cannot in principle factor out the dependence on
  $f(\vec{v})$ since it rests on assumptions on the time dependence of
  the modulation which depend on $f(\vec{v})$ itself.}, a general
strategy has been developed \cite{mccabe_eta,gondolo_eta} to factor
out the dependence on $f(\vec{v})$ of the expected WIMP--nucleus
differential rate $dR/d E_R$ at the given recoil energy $E_R$. This
approach exploits the fact that $dR/d E_R$ depends on $f(\vec{v})$
only through the minimal velocity $v_{min}$ that the WIMP must have to
deposit at least $E_R$, i.e.:

\begin{equation}
\frac{dR}{d E_R}\propto \eta(v_{min})\equiv \int_{|\vec{v}|>v_{min}}\frac{f(\vec{v})}{|\vec{v}|}\; d^3 v.
\label{eq:eta_tilde_ex}
\end{equation}

\noindent By mapping recoil energies $E_R$ into same ranges of
$v_{min}$ the dependence on $\eta(v_{min})$ and so on $f(\vec{v})$
cancels out in the ratio of expected rates on different targets,
provided that the kinematics of the process, and so the relation
between $E_R$ and $v_{min}$, is fixed. Specifically, a scenario that
extends the kinematics of the DM--nucleus scattering and that has been
proposed to alleviate the tension among different direct detection
experiments is Inelastic Dark Matter (IDM)\cite{inelastic}.  In this
class of models a DM particle $\chi$ of mass $m_{\chi}$ interacts with
atomic nuclei exclusively by up--scattering to a second state
$\chi^{\prime}$ with mass
$m_{\chi^{\prime}}=m_{\chi}+\delta$.  In the case of
exothermic Dark Matter \cite{exothermic} $\delta<0$ is also possible:
in this case the particle $\chi$ is metastable and down--scatters to a
lighter state $\chi^{\prime}$. The halo--model factorization approach,
which has been recently extended to the inelastic case in the analysis
of direct--detection data\cite{halo_independent_inelastic,noi}, is
significantly more complicated compared to the elastic case, because
when $\delta\ne$0 the mapping from $E_R$ to $v_{min}$ is no longer a
one--to--one correspondence.

As far as the scaling law (\ref{eq:scaling_law}) is concerned, its
main motivation is probably that it corresponds to the dominant term
in the Neutralino--nucleus cross section predicted in Supersymmetry.
A simple phenomenological generalization of Eq. (\ref{eq:scaling_law})
consists in the Isospin violation mechanism (Isospin Violating DM,
IVDM) \cite{isospin_violation}, where a specific choice of the ratio
$r\simeq f_n/f_p\simeq Z/(Z-A)$ can suppress the WIMP coupling to a
given target\footnote{It should be pointed out that Isospin Violation
  is also predicted in Neutralino--nucleus scattering, although its
  relevance is limited to a very tuned choice of the fundamental susy
  parameters.}. The presently most constraining experiment at light
WIMP masses ($m_{\chi}\lsim$ 20 GeV) uses Germanium (SuperCDMS) while
those most constraining at larger WIMP masses use Xenon (LUX,
XENON100). By tuning $r$ to either $r\simeq$ -0.78 to suppress the
WIMP coupling to Germanium or $r\simeq -0.69$ to do the same for Xenon
the tension between different experiments can be at least alleviated
for each of the two different WIMP mass ranges.  Specifically, the
presence of different isotopes limits in practice the maximal
achievable cancellation between different targets, as quantified by
the maximal relative degrading factors tabulated in Tab I of
Ref.\cite{isospin_violation}, and defined as the maximal factor by
which the ratio between the expected rates on two given targets can be
reduced compared to the isospin--conserving case.

Lately, several independent analyzes\cite{ge_phobic,noi} have single
out a specific scenario where the three candidate WIMP events claimed
by the CDMS--Si experiment \cite{cdms_si} can be reconciled to the
bounds from SuperCDMS \cite{super_cdms} and XENON100\cite{xenon100} by
advocating exothermic scattering (i.e. IDM with $\delta<0$) and
$r\simeq$ -0.78 (i.e. IVDM with suppression of the WIMP--Ge
coupling). This compatibility, which after the subsequent
LUX\cite{lux} experiment bound can now only be achieved if the
function $\eta(v_{min})$ is assumed to be different to that predicted
by the Isothermal Sphere Model, is limited to the ranges: 1 GeV $\lsim
m_{\chi}\lsim$ 4 GeV, -270 keV$\lsim \delta\lsim$ -40 keV \cite{noi}.

The above Ge-phobic exothermic DM scenario, albeit tuned, is also
potentially informative on the WIMP--nucleus interaction.  For this
reason several authors\cite{cdms_si_eff,cdms_si_eff2} have discussed
it using the effective Lagrangian of Eq. (\ref{eq:eff_lagrangian}).

Recently, Next-To-Leading Order (NLO) corrections to the WIMP--nucleus
cross section have been estimated using Chiral Perturbation Theory,
including two-nucleon amplitudes and recoil-energy dependent shifts to
the single-nucleon scalar form factors
\cite{isospin_violation_nlo,isospin_violation_nlo2}. While some of the
matrix elements needed to numerically evaluate such corrections are
only known for closed shells and a rough extrapolation is needed to
apply the formalism of \cite{isospin_violation_nlo,
  isospin_violation_nlo2} to the nuclei used in real--life
experiments, including $Ge$ and $Si$, NLO corrections lead to two
important qualitative changes in the scaling law of
Eq.(\ref{eq:scaling_law}): (i) the cancellation leading to the
suppression of the coupling of the WIMP particle to a given nucleus is
no longer between the two one--nucleon terms proportional to $f_p$ and
$f_n$, but between their sum and the new two--nucleon
contribution. This implies that the value of the ratio $r$ that
maximizes the degrading factor can be very different from the Leading
Order (LO) case (for instance, the "standard'' value $r$=-0.78 for
Ge--phobic DM can be shifted to values below -2 or even to positive
values). (ii) The degrading factor acquires new energy--dependent
terms so that the cancellation involved in the IVDM scenario is
further spoiled (besides the effect due to the presence of more than
one isotope) because it requires different values of the couplings
across the experimental ranges of the recoil energy.

In light of the elements listed above in the present paper we wish to
extend the analysis of the Ge-phobic exothermic DM scenario in several
directions:
\begin{itemize}
\item we fully incorporate the halo--independent approach by
  introducing an appropriate definition of compatibility ratio which
  extends the definition of the degrading factors introduced in
  Ref.\cite{isospin_violation};
\item we explore the coupling constant parameter space of the
  effective model of Eq. (\ref{eq:eff_lagrangian}) in order to discuss
  the maximal achievable degrading factors within the IVDM scenario as
  well as the minimal values of the suppression scale $M_*$ required
  to explain the three CDMS--Si events in terms of WIMP scatterings;
\item we wish to discuss the effect on such an analysis of the
  inclusion of the NLO corrections of
  Ref\cite{isospin_violation_nlo,isospin_violation_nlo2};
\item we include a discussion on the thermal relic density of the
  metastable state $\chi$, showing in which circumstances it can be
  compatible to observation;
\item we also discuss accelerator bounds by showing the Large Hadron
  Collider (LHC) constraints from monojet and hadronically-decaying
  mono-W/Z searches. The latter results need to assume that the
  validity of the effective theory of Eq.(\ref{eq:eff_lagrangian})
  extends to the LHC energy scale \cite{eft_validity_lhc}.
\end{itemize}

Our paper is organized as follows: in Section \ref{sec:model} we
summarize the effective model we use as well as the expressions
relevant to expected direct detection rates and the halo--independent
factorization; in Section \ref{sec:isospin_violation} we discuss
several aspect of the mechanism of isospin--violation, both at the
Leading Order and at the Next-to-Leading Order; in Section
\ref{sec:cdms-si} we discuss the CDMS-Si excess and its connection to
exothermal Ge-phobic DM; in Section \ref{sec:relic_abundance} a
discussion on the metastable state $\chi$ lifetime and its thermal
relic density is provided; in Section \ref{sec:LHC} we give the
details of our simulation for monojet and hadronically-decaying
mono-W/Z searches at the LHC; in Section \ref{sec:results} we combine
all the elements of the previous Sections to provide a quantitative
discussion of the phenomenology of our DM candidate; finally, our
Conclusions are contained in Section \ref{sec:conclusions}.

\section{The model}
\label{sec:model}
We generalize the Lagrangian of Eq.(\ref{eq:eff_lagrangian}) to an
inelastic coupling involving the two Dirac particles $\chi$ and
$\chi^{\prime}$ and slightly modify the ensuing formulas by
factorizing in each coupling the corresponding quark mass:

\begin{equation}
 {\cal L}=\sum_{q=u,d,s,c,b,t} \frac{m_q \tilde{\lambda}_q}{\Lambda^3}\bar{\chi}^{\prime}\chi \bar{q} q \mbox{+h.c.}
\label{eq:eff_lagrangian_full}
\end{equation}

\noindent 

Below the scale of the heavy quarks the latter can be
integrated out leading to the effective Lagrangian:

\begin{equation}
  {\cal L}=\sum_{q=u,d,s} \frac{m_q \lambda_q}{\Lambda^3}\bar{\chi}^{\prime}\chi \bar{q} q +
  \sum_{q=u,d,s} \frac{\lambda_{\theta}}{\Lambda^3}\bar{\chi}^{\prime}\chi \theta^{\mu}_{\mu} +\mbox{h.c.}
\label{eq:eff_lagrangian_integrated_out}
\end{equation}

\noindent where $\theta^{\mu}_{\mu}$ is the trace of the
stress--energy tensor,
$\lambda_{\theta}\equiv2/27\sum_{Q=c,b,t}\tilde{\lambda}_Q$ while
$\lambda_q \equiv \tilde{\lambda}_q-\lambda_{\theta}$.  The
phenomenology depends only on the ratios $\lambda_q/\Lambda^3$,
$\lambda_{\theta}/\Lambda^3$ so it is possible to absorb one among the
couplings, for definiteness $\lambda_u$, in the definition of the
suppression scale, i.e. $1/\tilde{\Lambda}^3\equiv
\lambda_u/\Lambda^3$ and normalize all the other couplings to
$\lambda_u$, i.e. $\bar{\lambda_q}\equiv \lambda_q/\lambda_u$,
$\bar{\lambda}_{\theta}\equiv \lambda_{\theta}/\lambda_u$. In this way
the effective lagrangian depends on four independent parameters.

The ensuing WIMP--nucleus scattering differential rate is given by the expression:

\begin{eqnarray}
\frac{d R}{d E_R}&=&MT\frac{\rho_{\chi} m_N}{2\tilde{\Lambda}^6 \pi m_{\chi}} N_T \sum_A f_A \left |Z f_p + (A-Z) f_n \right |^2 \times\nonumber\\
&&F(E_R)^2\eta(v_{min}(E_R)),
\label{eq:diff_rate}
\end{eqnarray}

\noindent where $m_N$ is the mass of the target nucleus, $M$ is the
detector mass, $T$ the time exposition, $\rho_{\chi}$ is the local
mass density of the $\chi$ particles in the neighborhood of the Sun,
$N_T$ is the number of targets per unit detector mass, $f_A$ is the
fractional abundance of nuclei with mass number $A$ in case more than
one isotope is present and $F(E_R)$ is a form factor taking into
account the finite size of the nucleus, for which we assume the
standard form\cite{helm}:

\begin{eqnarray}
  F(E_R)&=&\frac{3}{qR^{\prime}}\left [ \frac{sin(qR^{\prime})}{(qR^{\prime})^2} - \frac{cos(qR^{\prime})}{qR^{\prime}}  \right]\exp\left (-\frac{(qs)^2}{2} \right ) \nonumber\\
  q&=&\sqrt{2 m_N E_R};\;\; R^{\prime}=\sqrt{R_N^2-5 s^2}\nonumber\\
  R_N&=&1.2 A^{\frac{1}{3}};\;\;s=\mbox{1 fm},
\label{eq:form_factor}
\end{eqnarray}

while:

 \begin{equation}
 f_{p,n}=\frac{\sigma_{\pi N}}{m_u+m_d} \left [ m_u (1\pm\xi)+m_d \bar{\lambda}_d (1\mp\xi) \right ]+\bar{\lambda}_s \sigma_s+\bar{\lambda}_{\theta} m_p,
 \label{eq:f_p_f_n}
 \end{equation}

 \noindent with $\sigma_{\pi N}=((m_u+m_d)/2)<p|\bar{u}u+\bar{d}d|p>$,
 $\xi=<p|\bar{u}u-\bar{d}d|p>/<p|\bar{u}u+\bar{d}d|p>$,
 $\sigma_s=<p|m_s\bar{s}s|p>$ \footnote{In the analysis of Section
   \ref{sec:results} we will assume $\sigma_{\pi N}$=45 MeV,
   $\sigma_s$=45 MeV, $\xi$=0.18 \cite{nuclear_constants}.}. Finally
 the function $\eta$ parametrizes the dependence on the WIMP velocity
 distribution:

\begin{equation}
\eta(v_{min})=\int_{|\vec{v}|>v_{min}}\frac{f(\vec{v})}{|\vec{v}|}\; d^3 v,
\label{eq:eta}
\end{equation}

\noindent with:

\begin{equation}
v_{min}(E_R)=\frac{1}{\sqrt{2 m_N E_R}}\left | \frac{m_NE_R}{\mu_{\chi N}}+\delta \right |.
\label{eq:vmin}
\end{equation}

\noindent In the above equation $\mu_{\chi N}$ is the WIMP--nucleus reduced mass.

In a real--life experiment $E_R$ is obtained by measuring a related
detected energy $E^{\prime}$ obtained by calibrating the detector with
mono--energetic photons with known energy. However the detector
response to photons can be significantly different compared to the
same quantity for nuclear recoils.  For a given calibrating photon
energy the mean measured value of $E^{\prime}$ is usually referred to
as the electron--equivalent energy $E_{ee}$ and measured in keVee. On
the other hand $E_R$ (that represents the signal that would be
measured if the same amount of energy were deposited by a nuclear
recoil instead of a photon) is measured in keVnr. The two quantities
are related by a quenching factor $Q$ according to $E_{ee}=Q(E_R) E_R$
\footnote{In the following Sections we will focus on bolometric
  detectors (SuperCDMS, CDMS--$Si$) for which we will assume $Q$=1.}.
Moreover the measured $E^{\prime}$ is smeared out compared to $E_{ee}$
by the energy resolution (a Gaussian smearing
$Gauss(E_{ee}|E^{\prime},\sigma_{rms}(E^{\prime}))\equiv
1/(\sigma_{rms}\sqrt{2\pi})exp[-(E^{\prime}-E_{ee})^2/(2\sigma_{rms}^2)]$
with standard deviation $\sigma_{rms}(E^{\prime})$ related to the Full
Width Half Maximum (FWHM) of the calibration peaks at $E^{\prime}$ by
$FHWM=2.35 \sigma_{rms}$ is usually assumed) and experimental count
rates depend also on the counting efficiency or cut acceptance
$\epsilon(E^{\prime})$.  Overall, the expected differential event rate
is given by:

\begin{eqnarray}
  \frac{dR}{d E^{\prime}}&=&\epsilon(E^{\prime})\int_0^{\infty}d E_{ee} Gauss(E_{ee}|E^{\prime},\sigma_{rms}(E^{\prime}))\times\nonumber\\
&& \frac{1}{Q(E_R)} \frac{d R}{d E_R}.
\label{eq:rate_folding}
\end{eqnarray}

\subsection{Factorization of halo dependence}
\label{sec:model_factorization}

In the isospin--conserving case $f_n=f_p$ it is customary to factorize
in Eq.(\ref{eq:diff_rate}) the WIMP--proton point--like cross section,
$\sigma_p=\mu_{\chi p}f_p^2/(\tilde{\Lambda}^6\pi)$, with $\mu_{\chi
  p}$ the WIMP--proton reduced mass. In the isospin--violating case it
may be more convenient to factorize the WIMP--neutron cross section
instead (for instance in the case $f_p\ll f_n$) or, actually, any
other conventional cross section:

\begin{equation}
  \sigma_0 =\frac{\mu_{\chi
      p}^2f_0^2}{\tilde{\Lambda}^6\pi},
\label{eq:sigma_0}
\end{equation}

\noindent with $f_0$ an arbitrary amplitude. No matter what amplitude
is factorized, it is always possible to recast the differential rate
in the form:

\begin{equation}
\frac{dR}{dE_R}[E_R(v_{min})]=M T \frac{N_T m_N\tilde{A}^2}{2 \mu_{\chi p}^2} F^2(E_R) \tilde{\eta}(v_{min}),
\label{eq:dr_de_recast}
\end{equation}

\noindent with:

\begin{equation}
\tilde{A}=Z \frac{f_p}{f_0}+(A-Z)\frac{f_n}{f_0},
\label{eq:scaling_law_f0}
\end{equation}

\noindent and where the quantity:

\begin{equation}
\tilde{\eta}(v_{min})\equiv \frac{\rho_{\chi}}{m_{\chi}} \sigma_0 \eta(v_{min}),
\label{eq:eta_tilde}
\end{equation}

\noindent is a factor common to the WIMP--rate predictions of all
experiments, provided that it is sampled in the same intervals of
$v_{min}$. Mutual compatibility among different detectors' data can
then be investigated (factorizing out the dependence on the halo
velocity distribution) by binning all available data in the same set
of $v_{min}$ intervals and by comparing the ensuing estimations of
$\tilde{\eta}(v_{min})$.

Combining Eqs.(\ref{eq:dr_de_recast}) and (\ref{eq:rate_folding}) the
expected number of events in the interval
$E_1^{\prime}<E^{\prime}<E_2^{\prime}$ can be cast in the form:

\begin{eqnarray}
&& \bar{R}(E_1^{\prime},E_2^{\prime})=\int_{E_1^{\prime}}^{E_2^{\prime}} d E^{\prime} \frac{dR}{d E^{\prime}}=\nonumber\\
&&=\int_{0}^{\infty} d E_{ee} \tilde{\eta}\left \{v_{min}\left [E_R\left (E_{ee} \right )  \right ] \right \} {\cal R}_{[E_1^{\prime},E_2^{\prime}]}(E_{ee}),
\end{eqnarray}

\noindent where the response function ${\cal R}$, given by:

\begin{eqnarray}
&&{\cal R}_{[E_1^{\prime},E_2^{\prime}]}(E_{ee})=\frac{N_T m_N \tilde{A}^2}{2 \mu_{\chi p}^2}F^2\left[E_R(E_{ee}) \right ] MT\times\nonumber\\
&&\int_{E_1^{\prime}}^{E_2^{\prime}} d E^{\prime} Gauss(E_{ee}|E^{\prime},\sigma_{rms}(E^{\prime})) \epsilon(E^{\prime}),
\label{eq:response_function}
\end{eqnarray}

\noindent contains the information of each experimental
setup. Given an experiment with detected count rate $N_{exp}$ in the
energy interval $E_1^{\prime}<E^{\prime}<E_2^{\prime}$ the combination:

\begin{eqnarray}
  \bar{\tilde{\eta}}&=&\frac{\int_{0}^{\infty} d E_{ee} \tilde{\eta}(E_{ee}) {\cal R}_{[E_1^{\prime},E_2^{\prime}]}(E_{ee})}
  {\int_{0}^{\infty} d E_{ee} {\cal R}_{[E_1^{\prime},E_2^{\prime}]}(E_{ee})}\nonumber\\
&=&\frac{N_{exp}}{\int_{0}^{\infty} d E_{ee} {\cal R}_{[E_1^{\prime},E_2^{\prime}]}(E_{ee})},
\label{eq:eta_bar_e}
\end{eqnarray}

\noindent can be cast in the form\cite{gondolo_eta}:

\begin{eqnarray}
  \bar{\tilde{\eta}}&=&\frac{\int_{0}^{\infty} d v_{min} \tilde{\eta}(v_{min}) {\cal R}_{[E_1^{\prime},E_2^{\prime}]}(v_{min})}
  {\int_{0}^{\infty} d v_{min} {\cal R}_{[E_1^{\prime},E_2^{\prime}]}(v_{min})}\nonumber\\
&=&\frac{N_{exp}}{\int_{0}^{\infty} d v_{min} {\cal R}_{[E_1^{\prime},E_2^{\prime}]}(v_{min})},
\label{eq:eta_bar_vmin}
\end{eqnarray}

\noindent by changing variable from $E_{ee}$ to $v_{min}$ (in the
above expression ${\cal R}_{[E_1^{\prime},E_2^{\prime}]}(v_{min})$ =
${\cal R}_{[E_1^{\prime},E_2^{\prime}]}(E_{ee})$ $d E_{ee}/d v_{min}$)
and can be interpreted as an average of the function
$\tilde{\eta}(v_{min})$ in an interval
$v_{min,1}<v_{min}<v_{min,2}$. The latter is defined
as the one where the response function ${\cal R}$ is ``sizeably''
different from zero (we will conventionally take the interval
$v_{min}[E_R(E_{ee,1})]<v_{min}<v_{min}[E_R(E_{ee,2})]$ with
$E_{ee,1}=E^{\prime}_1-\sigma_{rms}(E^{\prime}_1)$,
$E_{ee,2}=E^{\prime}_2+\sigma_{rms}(E^{\prime}_2)$, i.e. the
$E^{\prime}$ interval enlarged by the energy resolution).

A complication of the IDM case (compared to elastic scattering) is that
the mapping between $v_{min}$ and $E_R$ (and so $E^{\prime}$) from
Eq. (\ref{eq:vmin}) is no longer univocal. In particular
$v_{min}$ has a minimum when $E_R$=$E_R^*$=$|\delta|\mu_{\chi N}/m_N$ given by:

\begin{equation}
v_{min}^*=\left \{ \begin{array}{ll}
\sqrt{\frac{2|\delta|}{\mu_{\chi N}}} & \mbox{if $\delta>0$ }\\
0 &   \mbox{if $\delta<0$},
\end{array}
  \right .
\label{eq:vstar}
\end{equation}

\noindent and any interval of $v_{min}>v_{min}^*$ corresponds to two
mirror intervals for $E_R$ with $E_R<E_R^*$ or $E_R>E_R^*$.  As a
consequence of this when $E_{ee}(E^*_R)\in [E_{ee,1},E_{ee,2}]$ the
change of variable from Eq.(\ref{eq:eta_bar_e}) to
Eq.(\ref{eq:eta_bar_vmin}) leads to two disconnected integration
ranges for $v_{min}$ and to an expression of $\bar{R}$ in terms of a
linear combination of the corresponding two determinations of
$\bar{\tilde{\eta}}$. This problem can be easily solved by binning the
energy intervals in such a way that for each experiment the energy
corresponding to $E_{ee}(E_R^*)$ is one of the bin
boundaries\cite{noi}.

\section{Isospin violation}
\label{sec:isospin_violation}

The differential rate (\ref{eq:dr_de_recast}) depends on the couplings $\bar{\lambda}_d$, $\bar{\lambda}_s$ and $\bar{\lambda}_{\theta}$ and
on the suppression scale $\tilde{\Lambda}$
 only through the cross section $\sigma_0$ and the scaling law $\tilde{A}$.
 Following \cite{isospin_violation} a degrading factor can be introduced, as the ratio of the expected rate, for some value of $r$, normalized to
 $r=1$:

 \begin{equation}
 D(r,\sigma_0)\equiv \frac{\bar{R}(r,\sigma_0)}{\bar{R}(r=1,\sigma_0)}=\frac{\bar{R}(r)}{\bar{R}(r=1)},
 \label{eq:degrading_factor}
 \end{equation}

 \noindent where the dependence on $\sigma_0$ and so on the
 suppression scale $\tilde{\Lambda}$ cancels out in the ratio.  The
 degrading factor is minimized if in Eq.(\ref{eq:scaling_law_f0})
 $r\equiv f_n/f_p=r_{min}\simeq Z/(Z-\bar{A)}$ where $\bar{A}$ is some
 average of the atomic mass numbers over the isotopical abundances. On
 the other hand, for a fixed value of $\bar{\lambda}_{\theta}$ and
 $\bar{\lambda}_s$, setting $r=r_{min}$ corresponds through
 Eq.(\ref{eq:f_p_f_n}) to fixing $\bar{\lambda}_d$ to some value
 $\bar{\lambda}_{d,min}$. Notice that while $r_{min}$ is fixed to a
 single value, $\bar{\lambda}_d$ depends on $\bar{\lambda}_{\theta}$
 and $\bar{\lambda}_s$. In order to discuss the relic abundance and
 the signals at the LHC the suppression scale $\tilde{\Lambda }$ must
 be fixed (we will do that by requiring that the expected number of
 events can explain the CDMS--$Si$ excess) as well as each of the
 heavy--quark couplings $\bar{\lambda}_{Q=c,b,t}$. As far as the
 latter are concerned, only their sum is determined through
 $\bar{\lambda}_{\theta}$. In Sections \ref{sec:relic_abundance} and
 \ref{sec:LHC} we will choose to fix them with the goal to minimize
 the $\chi$ thermal relic abundance.  Then, following
 \cite{isospin_violation_nlo2} we will perform our phenomenological
 discussion into the plane
 $\bar{\lambda}_{\theta}$--$\bar{\lambda}_s$.

\subsection{Leading-order result}
\label{sec:lo}

It is now instructive to rewrite explicitly the scaling law in Eq.(\ref{eq:scaling_law_f0}) in terms of the couplings $\bar{\lambda}_q$:

\begin{eqnarray}
f_0 \tilde{A}&=&Z\sum_{i=u,d,s,\theta} M_i \bar{\lambda}_i+A \sum_{i=u,d,s,\theta} P_i \bar{\lambda}_i\nonumber\\
&=&Z\left [ M_d \bar{\lambda}_d + M(\bar{\lambda}_s,\bar{\lambda}_{\theta}) \right ]\nonumber\\
&+&A \left [ P_d \bar{\lambda}_d + P(\bar{\lambda}_s,\bar{\lambda}_{\theta})\right ],
\label{eq:scaling_law_lambda_q}
\end{eqnarray}

\noindent where $M(\bar{\lambda}_s,\lambda_{\theta})\equiv M_u+M_s \bar{\lambda}_s +M_{\theta} \bar{\lambda}_{\theta}$ and
$P(\bar{\lambda}_s,\lambda_{\theta})\equiv P_u+P_s \bar{\lambda}_s +P_{\theta} \bar{\lambda}_{\theta}$ and
the explicit expressions of the coefficients $M_i$ and $P_i$ are given for convenience in Table \ref{table:p_m_coefficients}.

\begingroup
\squeezetable
\begin{table}[t]
{\begin{tabular}{@{}|c|c|c|c|c|@{}}
\hline
    &  $u$  & $d$  & $s$  & $\theta$  \\
\hline
$P_q$ &  $\frac{\sigma_{\pi N}}{m_u+m_d}m_u (1-\xi) [+t_u] $& $\frac{\sigma_{\pi N}}{m_u+m_d}m_d(1+\xi) [+t_d]$ & $ \sigma_s[+t_s]$ & $ m_p$  \\
$M_q$ &  2 $\frac{\sigma_{\pi N}}{m_u+m_d}m_u \xi $  & -2 $\frac{ \sigma_{\pi N}}{m_u+m_d}m_d \xi $ & 0 &  $0$  \\
\hline
\end{tabular}}
\caption{Coefficients entering the expression of the scaling law of
  Eq.(\ref{eq:scaling_law_lambda_q}). In parenthesis are given the
  additional terms (whose numerical values are given in Table \ref{table:nlo}) to be used in Eq.(\ref{eq:scaling_law_lambda_q}) when
  the approximate NLO--corrected expression of
  Eq.(\ref{eq:diff_rate_nlo_approx}) is adopted. 
  \label{table:p_m_coefficients}}
\end{table}
\endgroup

At fixed $\bar{\lambda}_{\theta}$ and $\bar{\lambda}_s$ the scaling
law and the degrading factor are minimized as a function of
$\bar{\lambda}_d$ (or, equivalently of $r$, since there is a
one-to-one correspondence between them through
Eq.(\ref{eq:f_p_f_n})). There is however a particular choice of
$\bar{\lambda}_{\theta}$ and $\bar{\lambda}_s$ such that:

\begin{equation}
  \frac{M(\bar{\lambda}_s,\bar{\lambda}_{\theta})}{M_d}=\frac{P(\bar{\lambda}_s,\bar{\lambda}_{\theta})}{P_d}\equiv f(\bar{\lambda}_s,\bar{\lambda}_{\theta}).
\label{eq:allignement}
\end{equation}

\noindent In this particular case the scaling law acquires the
factorization:
\begin{equation}
f_0\tilde{A}=\left [ \bar{\lambda}_d+f(\bar{\lambda}_s,\bar{\lambda}_{\theta}) \right ]\left [M_d Z + P_d A \right ],
\label{eq:scaling_law_factorized}
\end{equation}

\noindent and the minimum of the degrading factor is obtained for
$\bar{\lambda}_{d,min}=-f(\bar{\lambda}_s,\bar{\lambda}_{\theta})$.
This result may seem puzzling because $\bar{\lambda}_{d,min}$ in this
case is the same for all nuclei, i.e. this special {\it alignment} of
the coupling constants corresponds to a vanishing signal for all
nuclei. Another way to see this is that when Eq.(\ref{eq:allignement})
is satisfied the values of $r_{min}\simeq Z/(Z-\bar{A)}$ of different
nuclei are mapped into the same $\bar{\lambda}_{d,min}$.  It is
trivial to verify that this situation simply corresponds to a
vanishing $f_p$ {\it at fixed $r$}, so that $f_n=r f_p\rightarrow 0$:
physically, the WIMP cross sections on protons and neutrons are both
vanishing.  This implies an overall rescaling of all the signals on
different targets, but since this is done at fixed $r$, the relative
degrading factors among different nuclei can be fixed to those
required by the isospin--violation scenario in order to allow
compatibility among signals and constraints.  In particular, the
condition (\ref{eq:allignement}) implies:

\begin{equation}
m_s \bar{\lambda}_s +m_p \bar{\lambda}_{\theta}+\frac{2 m_u}{m_u+m_d}\sigma_{\pi,N}=0.
\label{eq:straight_line}
\end{equation}

\noindent The factorization of Eq.(\ref{eq:scaling_law_factorized})
has also another important feature: when a specific nuclear target is
considered the scaling law converges to the same value for any choice
of the mass number $A$, so that the degrading factor of
Eq.(\ref{eq:degrading_factor}) can in principle become arbitrarily
small even in presence of many isotopes. As a consequence of this, the
parameter space close to the straight line of
Eq.(\ref{eq:straight_line}) in the plane
$\bar{\lambda}_{\theta}$--$\bar{\lambda}_s$ corresponds to a situation
where the scale $\tilde{\Lambda}$ can be maximally suppressed at fixed
$\bar{R}(E_1^{\prime},E_2^{\prime})$, i.e. {\it if IVDM is advocated
  to explain a given experimental excess, $\tilde{\Lambda}$ reaches
  its minimum values when $\bar{\lambda}_{\theta}$ and
  $\bar{\lambda}_s$ are close to the straight line of
  Eq.(\ref{eq:straight_line})}.

\subsection{Next-to-Leading order corrections}
\label{sec:nlo}

Recently, NLO corrections to WIMP--nucleus elastic scattering have been estimated using Chiral Perturbation Theory\cite{isospin_violation_nlo,isospin_violation_nlo2}.
In the following we will adopt the same corrections also for the inelastic case, since
in the interaction induced by Eq.(\ref{eq:eff_lagrangian_full}) the terms which depend on the nuclear state
are factorized from those depending on the WIMP states. Moreover, the mass splitting $\delta$ is much smaller than
any scale in the nucleus. Including NLO corrections Eq.(\ref{eq:diff_rate}) is modified as\cite{isospin_violation_nlo2}:

\begin{eqnarray}
&&\frac{d R}{d E_R}=MT\frac{\rho_{\chi} m_N}{2\tilde{\Lambda}^6 \pi m_{\chi}} N_T \times \nonumber\\
&&\sum_A f_A \left |(Z f^{NLO}_p + (A-Z) f^{NLO}_n)F(E_R)+A f^{NLO}_{2N} \right |^2 \times\nonumber\\
&&\eta(v_{min}(E_R)),
\label{eq:diff_rate_nlo}
\end{eqnarray}

\noindent where:

\begin{eqnarray}
f^{NLO}_p &=& f_p+A E_R \sum_{q=u,d,s}\bar{\lambda}_q s^q_p,\nonumber\\
f^{NLO}_n &=& f_n+A E_R \sum_{q=u,d,s}\bar{\lambda}_q s^q_n,\nonumber\\
f^{NLO}_{2N} &=& (t_u+\bar{\lambda}_d t_d) F_{\pi\pi}(E_R)+\bar{\lambda}_s t_s F_{\eta\eta}(E_R).
\label{eq:nlo_amplitudes}
\end{eqnarray}

\noindent The numerical values of the coefficients $s^q_{p,n}$ and $t_q$ are given for convenience in Table \ref{table:nlo} and,
 along with  the form factors:

\begin{eqnarray}
F_{\pi\pi}(E_R)&=& F_{exp}(|q|) \left [ (1.20 -1.83 A^{-\frac{1}{3}}+4.60 A^{-\frac{2}{3}})|q| \right ],\nonumber\\
F_{\eta\eta}(E_R)&=& F(E_R) \left [ (0.74 +1.04 A^{-\frac{1}{3}}-1.93 A^{-\frac{2}{3}})|q| \right ],\nonumber\\
F_{exp}(|q|)&=& \exp(-|q|^2 R_0^2/6)
\label{eq:form_factors_nlo}
\end{eqnarray}

\noindent with 
$|q|=\sqrt{2 m_A E_R}$ and $R_0=[0.3+0.91
(m_A/\mbox{GeV})^{\frac{1}{3}}$ are taken from
\cite{isospin_violation_nlo2} and subject to large uncertainties (in
the equation above $F(E_R)$ is the same of
Eq.(\ref{eq:form_factor})). Specifically, they are only known for
closed-shell nuclei and strictly speaking could not be used for nuclei
such as Silicon or Germanium. However we wish here to discuss some
qualitative properties that descend from the modified scaling law of
Eq.(\ref{eq:diff_rate_nlo}) and that depend only mildly on the actual
values of the parameters.

\begingroup
\squeezetable
\begin{table}[t]
\begin{center}
{\begin{tabular}{@{}|c|c|c|c|c|c|c|c|@{}}
\hline
  $s_p^u$  & $s_p^d$  & $s_n^u$  & $s_n^d$  & $s_{p,n}^s$ & $t_u$ & $t_d$ & $t_s$ \\
\hline
-0.116 & -0.192  & -0.096 & -0.232 & -0.472 & -0.63 MeV & -1.27 MeV & 0.070 MeV  \\
\hline
\end{tabular}}
\caption{Coefficients entering the NLO amplitudes of Eq.(\ref{eq:nlo_amplitudes}) (from \cite{isospin_violation_nlo2}).
\label{table:nlo}}
\end{center}
\end{table}
\endgroup

The main qualitative conclusion of the analysis of
Ref. \cite{isospin_violation_nlo2} is that, in the NLO--corrected
differential rate of Eq.(\ref{eq:diff_rate_nlo}), the cancellation
mechanism at work in the IVDM scenario is different from the LO case,
namely it is no longer between the WIMP--proton amplitude $f^{NLO}_p$
and the WIMP--neutron amplitude $f^{NLO}_n$, but between the
combination of the latter and the new two--nucleon amplitude
$f^{NLO}_{2N}$. As a consequence of this the value of the ratio
$r=f^{NLO}_n/f^{NLO}_p$ corresponding to the minimum of the degrading
factor (\ref{eq:degrading_factor}) can be very different from the LO
case (for instance, in the case of Germanium, $r_{min}$ can be smaller
than -2 or even larger than zero, compared to the LO value
$r_{min}\simeq$-0.78). Another important difference with the LO case
is that in the NLO--corrected amplitude of
Eq. (\ref{eq:diff_rate_nlo}) it is no longer possible to factorize a
WIMP--nucleon cross section either off protons or neutrons. Moreover,
the NLO corrections of Eq.(\ref{eq:nlo_amplitudes}) include terms with
explicit dependence on the recoil energy, so that in the differential
rate of Eq.(\ref{eq:diff_rate_nlo}) the modified scaling law can no
longer be factorized and depends on the energy bin.  We wish now to
show that, in spite of all these apparently significant changes, the
phenomenology is only expected to change mildly with the exception of
specific situations.

In order to do this we start by noting that in
Eq.(\ref{eq:nlo_amplitudes}) the small numerical factors
$s^{u,d,s}_{p,n}$ are multiplied by recoil energies which are of order
keV, while the natural scale of the amplitudes $f_{p,n}$ is set by the
dimensional constants $\sigma_{\pi N}$, $\sigma_s$ and $m_p$, which
are all of order MeV, or even GeV (see Eq.(\ref{eq:f_p_f_n})). So,
with the exception of strong cancellations in the LO amplitudes
$f_{p,n}$, the energy--dependent terms can be safely neglected.
Notice that in Section \ref{sec:lo} the peculiar region of the
($\bar{\lambda}_{\theta}$--$\bar{\lambda}_s$) parameter space where
$f_{p,n}\rightarrow 0$ was already discussed and shown to be close to
the straight line of Eq.(\ref{eq:straight_line}). Clearly, in that
specific regime the energy--dependent terms in
Eq.(\ref{eq:nlo_amplitudes}) are no longer negligible, as we will
check explicitly in Section \ref{sec:results}. Another energy
dependence in the NLO corrections is contained in the form factors of
the the two--nucleon amplitude $f^{NLO}_{2N}$. As far as
$F_{\eta\eta}(E_R)$ is concerned, it is multiplied by the small factor
$t_s\ll t_{u,d}$ and can be neglected. On the other hand in the
specific case of a light WIMP that we will discuss in the following,
it will be safe to assume $F(E_R)$=$F_{\pi\pi}(E_R)$=1. With these
assumptions the rate of Eq.(\ref{eq:diff_rate_nlo}) can be
approximated by:

\begin{eqnarray}
&&\frac{d R}{d E_R}=MT\frac{\rho_{\chi} m_N}{2\tilde{\Lambda}^6 \pi m_{\chi}} N_T \times \nonumber\\
&&\sum_A f_A \left |(Z f_p + (A-Z) f_n)+A (t_u+t_d \bar{\lambda}_d) \right |^2 \eta(v_{min}(E_R))\nonumber \\
&&=M T \frac{N_T m_N\tilde{A}^2}{2 \mu_{\chi p}^2}\tilde{\eta}(v_{min}).
\label{eq:diff_rate_nlo_approx}
\end{eqnarray}

\noindent Notice that even in the approximate form above the
WIMP--proton cross section can no longer be factorized in the
differential rate.  So in the last step we have recast the
differential rate in a form suitable for a halo--independent analysis
by factorizing an (arbitrary) amplitude $f_0$ so that $\tilde{\eta}$
is defined by Eq.(\ref{eq:eta_tilde}) with, as usual, $\sigma_0
=\mu_{\chi p}f_0^2/(\tilde{\Lambda}^6\pi)$, while the scaling law is
explicitly given by:

\begin{equation}
f_0\tilde{A}=Z f_p+(A-Z) f_n +A (t_u+t_d \bar{\lambda}_d).
\label{eq:scaling_law_nlo}
\end{equation}

\noindent The expression above has clearly a different dependence on
the $r=f_n/f_p$ parameter compared to the LO scaling, so that the
$r_{min}$ values corresponding to the minimum of the NLO degrading
factor can be very different from the LO
case\cite{isospin_violation_nlo2}. However, this happens because $r$
is not suitable to parametrize both the LO and the NLO scaling
laws. On the other hand, in both cases the scaling law can be cast in
the form:

\begin{equation}
f_0\tilde{A} \propto A + Z t,
\label{eq:scaling_law_t}
\end{equation}

\noindent with $t=f_p/f_n-1$ in the LO case and
$t=(f_p-f_n)/(f_n+t_u+t_d \bar{\lambda}_d)$ in the (approximate) NLO
case.  Irrespective to the relation between the $t$ parameter and the
coupling constants, which is different in the LO and NLO cases, the
couplings enter in the degrading factor only though $t$, so in both
cases $t_{min}\simeq -\bar{A}/Z$ (with $\bar{A}$ some average of the
target atomic number over isotopes) and, most importantly, the minimum
values of the degrading factor defined in
Eq.(\ref{eq:degrading_factor}) and seen as a function of $t$ instead
of $r$ are the same in the LO and in the NLO cases. As a consequence
of this, as long as the approximation of
Eq.(\ref{eq:diff_rate_nlo_approx}) is valid, the phenomenology for
$t=t_{min}$ is not going to change as far as direct detection is
concerned from the LO to the NLO case, although, due to the different
mappings between the $t$ parameter and the
$\bar{\lambda}_q$,$\bar{\lambda}_{\theta}$ couplings, this can alter
the correlation with other types of signal.

Moreover, the NLO--corrected scaling law
(\ref{eq:scaling_law_nlo}) maintains the same form of
Eq.(\ref{eq:scaling_law_lambda_q}) as expressed as a function of the
couplings $\bar{\lambda}_{q,\theta}$ (where the coefficients are
modified by the terms shown in parenthesis in Table
\ref{table:p_m_coefficients}), so that the factorization of
Eq.(\ref{eq:scaling_law_factorized}) is still possible in the
parameter space where condition (\ref{eq:allignement}) is verified.
As in the LO case, also in the NLO one this happens along a straight
line in the ($\bar{\lambda}_{\theta}$--$\bar{\lambda}_s$) plane now
given by:

\begin{equation}
\sigma_s \bar{\lambda}_s +m_p \bar{\lambda}_{\theta}+\frac{2 m_u}{m_u+m_d}\sigma_{\pi,N}+\frac{m_u}{m_d}t_d+t_u=0.
\label{eq:straight_line_nlo}
\end{equation}

\noindent Notice that the situation is very similar to the case
discussed in Section \ref{sec:lo}, i.e. close to the line
(\ref{eq:straight_line_nlo}) the special combination of coupling
constants
$\bar{\lambda}_{d,min}=-f(\bar{\lambda}_s,\bar{\lambda}_{\theta})$
{\it leads to WIMP-decoupling from all nuclei at the same
  time}. Contrary to the LO case, however, this effect does not have
the trivial explanation that the WIMP--nucleon cross section vanishes
(i.e. $f_n,f_p\rightarrow$0 keeping fixed their ratio $r$): in this
case the cancellation involves also the two--nucleon
amplitude. Nevertheless, for practical purposes, the phenomenology is
only slightly modified compared to the LO case: in the
($\bar{\lambda}_{\theta}$--$\bar{\lambda}_s$) parameter space close to
the line of Eq.(\ref{eq:straight_line_nlo}) the scale
$\tilde{\Lambda}$ is suppressed at fixed scattering rate
$\bar{R}(E_1^{\prime},E_2^{\prime})$. Notice that the straight line of
Eq.(\ref{eq:straight_line_nlo}) is only slightly shifted compared to
that of Eq.(\ref{eq:straight_line}).

\section{Exothermic Ge--phobic dark matter and the CDMS--$Si$ excess}
\label{sec:cdms-si}

\begin{figure}[ht]
\begin{center}
\includegraphics[width=1\columnwidth,bb= 46 194 506 635]{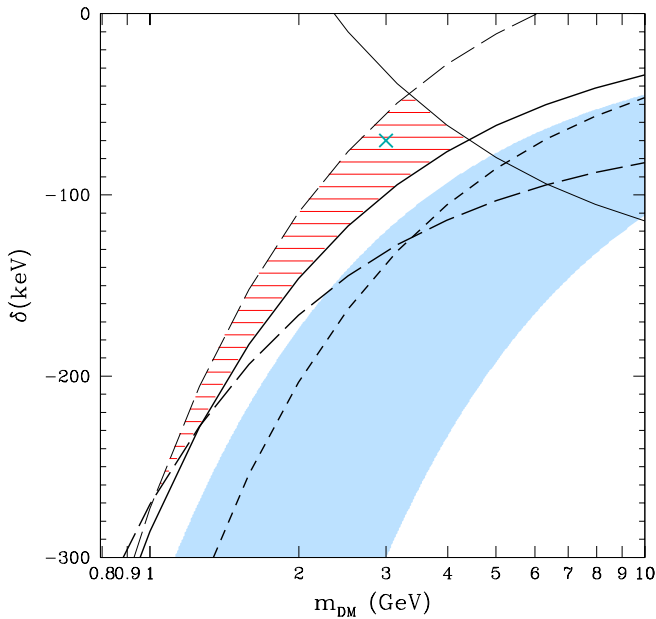}
\end{center}
\caption{ Mass splitting $\delta=m_{\chi^{\prime}}-m_{\chi}$ as a
  function of $m_{\chi}$.  The horizontally (red) hatched area
  represents the IDM parameter space where the excess measured by
  CDMS-$Si$\cite{cdms_si} corresponds to a $v_{min}<v_{esc}^{Lab}$
  range which is always below the corresponding one probed by LUX and
  XENON100 (adapted from \protect\cite{noi}). As explained in the
  text, in this case Xenon experiments can constrain the CDMS-$Si$
  excess only when some assumptions are made on the galactic velocity
  distribution. The enclosed region is the result of the combination
  of four conditions: the thin solid line where
  $v_{min}(E^{LUX}_{min})$=$v_{min}(E^{CDMS-Si}_{max})$; the thick
  solid line where
  $v_{min}(E^{LUX}_{min})$=$v_{min}(E^{CDMS-Si}_{min})$; the thin
  long--dashed line where
  $v_{min}(E^{CDMS-Si}_{max})$=$v_{esc}^{Lab}$; the thick long--dashed
  line where $v_{min}(E^{CDMS-Si}_{min})$=$v_{esc}^{Lab}$ (see
  text). The corresponding boundaries for XENON100 are less
  constraining: in particular the thin short--dashed line represents
  the parameter space where
  $v_{min}(E^{XENON100}_{min})$=$v_{min}(E^{CDMS-Si}_{min})$. The blue
  shaded strip represents points excluded by the consistency test
  introduced in Section 4.1 of Ref. \cite{noi}. The cross represents
  the benchmark point discussed in detail in Section
  \ref{sec:results}.}
\label{fig:mchi_delta_si}
\end{figure}

The CDM-$Si$ experiment\cite{cdms_si} has observed three WIMP
candidate events at energies $E_R$=8.2 keVnr, 9.5 keVnr and 12.3 keVnr
analyzing the full energy range 7 keVnr$<E_R<$100 keVnr with an
exposure of 140.2 kg day with a Silicon target. The probability
estimated by the same Collaboration that the known backgrounds would
produce three or more events in the signal region is 5.4\%.

An explanation of the three events observed by CDMS--$Si$ in terms of
a WIMP with a scalar isospin--conserving interaction (i.e. the scaling
law of Eq. (\ref{eq:scaling_law})) and assuming an isothermal sphere
model for the WIMP velocity distribution in the halo of our Galaxy
appears to be in strong disagreement with constraints from at least
three experiments: LUX\cite{lux}, XENON100\cite{xenon100} and
SuperCDMS \cite{super_cdms}. As shown by several authors
\cite{ge_phobic} for a light WIMP mass ($m_{\chi}\lsim$ 10 GeV) and an
exothermic interaction ($\delta<0$) the XENON100 constraint could be
evaded, while, in order to reconcile the result with the SuperCDSM
bound, Isospin Violation suppressing the WIMP coupling with Germanium
targets ($Ge$--phopic exothermic DM) could be advocated. This scenario
was however excluded by the subsequent LUX bound with its lower
threshold and higher exposure compared to XENON100, if an isothermal
sphere is assumed for the WIMP velocity distribution $f(\vec{v})$ in
our Galaxy; a halo--independent analysis, however, shows that the
CDMS--$Si$ excess and LUX can be compatible \cite{noi} provided that
the isothermal sphere assumption is abandoned and only minimal
assumptions are made on $f(\vec{v})$.

Namely, as discussed in detail in \cite{noi}, when no model is assumed
for the velocity distribution, an experiment can conservatively
constrain an excess claimed by another one only if it is sensitive to
the same $v_{min}$ interval, or to {\it lower} values. The latter
condition descends from the minimal requirement that the
$\tilde{\eta}$ function defined in (\ref{eq:eta},\ref{eq:eta_tilde})
is decreasing monotonically with $v_{min}$ (since $v_{min}$ is the
lower bound of an integration of the positive function
$f(\vec{v})/|\vec{v}|$). At the same time the requirement that the
WIMPs are gravitationally bound to our Galaxy implies that the signal
region must verify the condition that $v_{min}<v_{esc}^{Lab}$, with
$v_{esc}^{Lab}$ the Galactic escape velocity boosted in the Lab rest
frame. The two latter conditions depend on the mapping between $E_R$
and $v_{min}$, which, according to Eq.(\ref{eq:vmin}), in the IDM
scenario can be modified by assuming $\delta\ne$0. In particular, if,
for the same choice of $m_{\chi}$ and $\delta$, conflicting
experimental results can be mapped into non--overlapping ranges of
$v_{min}$ and if the $v_{min}$ range of the constraint is at higher
values compared to the excess (while that of the signal remains below
$v_{esc}^{Lab}$) the tension between the two results can be eliminated
by an appropriate choice of the $\tilde{\eta}$ function. This at the
price of having to assume that the function $\tilde{\eta}$ drops to
appropriately low values in the (high) $v_{min}$ range pertaining to
the constraint.

The result of a similar analysis in the $m_{\chi}$--$\delta$ parameter
space is shown in Fig. \ref{fig:mchi_delta_si}\cite{noi}.  In the case
of LUX we have assumed the range 2 PE$\le S_1\le$30 PE for the primary
scintillation signal $S_1$ (directly in Photo Electrons, PE) while for
XENON100 we have taken the experimental range 3 PE$\le S_1\le$30
PE. In both cases, following Eqs. (14-15) of
Ref. \cite{xenon100_response} we have modeled the detector's response
with a Poissonian fluctuation of the $S_1$ scintillation
photoelectrons combined with a Gaussian resolution $\sigma_{PMT}$=0.5
PE for the photomultiplier.  In Fig. \ref{fig:mchi_delta_si} the
horizontally (red) hatched area represents the IDM parameter space
where the excess measured by CDMS-$Si$\cite{cdms_si} corresponds to a
$v_{min}<v_{esc}^{Lab}$ range which is always below the corresponding
one probed by LUX and XENON100. In Fig. \ref{fig:mchi_delta_si} we
take $v_{esc}^{Lab}=$782 km/s (by combining the reference value of
the escape velocity $v_{esc}^{Galaxy}$=550 km/s in the galactic rest
frame with the velocity $v_0$=232 km/s of the Solar system with
respect to the WIMP halo) and the CDMS--$Si$ signal region is
approximated with the energy range 8 keVnr$<E_R<$12.5 keVnr.

From direct inspection of Fig.\ref{fig:mchi_delta_si}, indeed
exothermic DM (i.e. -260 keV$\lsim \delta\lsim$ -40 keV) is required
to allow compatibility between the CDM--Si excess and bounds from
liquid--Xenon scintillators, as well as very low values of the WIMP
particle mass ($m_{\chi}\lsim$ 4 GeV). Indicating the Region of
Interest of each experiment by $[E_{min},E_{max}]$, in
Fig.\ref{fig:mchi_delta_si} several curves are shown: on the thin
solid line $v_{min}(E^{LUX}_{min})$=$v_{min}(E^{CDMS-Si}_{max})$; on
the thick solid line
$v_{min}(E^{LUX}_{min})$=$v_{min}(E^{CDMS-Si}_{min})$; the thin
long--dashed line corresponds to
$v_{min}(E^{CDMS-Si}_{max})$=$v_{esc}^{Lab}$; the thick long--dashed
line to $v_{min}(E^{CDMS-Si}_{min})$=$v_{esc}^{Lab}$. The
corresponding boundaries for XENON100 are less constraining: in
particular the thin short--dashed line represents the parameter space
where $v_{min}(E^{XENON100}_{min})$=$v_{min}(E^{CDMS-Si}_{min})$. The
blue shaded strip represents points excluded by the consistency test
introduced in Section 4.1 of Ref. \cite{noi}, and does not affect the
compatibility region.

The same check can be made between the CDMS--$Si$ excess and the
SuperCDMS experiment bound\cite{super_cdms}, but in this case no
analogous compatibility region can be found in all the
$m_{\chi}$--$\delta$ plane. This means that, besides experimental
issues, the two measurements cannot be reconciled using kinematics
arguments only. However, in presence of some additional dynamical
mechanism suppressing WIMP scatterings on Germanium compared to that
on Silicon, the CDMS--$Si$ result and the SuperCDMS bound can in
principle be reconciled. An example of such mechanism is the
"$Ge$--phobic" IVDM scenario which is the main subject of
the present analysis.

  In Section \ref{sec:results} we will adopt the benchmark point
  represented by a cross in Fig.\ref{fig:mchi_delta_si} to perform a
  full numerical analysis of the IVDM parameter space.

\subsection{From direct--detection data to suppression scale}
\label{sec:lambda_from_data}

Following the halo--independent procedure outlined in Section
\ref{sec:model_factorization} it is straightforward, for a given
choice of the DM parameters, to obtain estimations
$\bar{\tilde{\eta}}_i^{CDMS-Si}$ of the function
$\tilde{\eta}(v_{min})$ averaged in appropriately chosen $v_{min}$
intervals mapped from the CDMS--Si Region of Interest (see
Eq.(\ref{eq:eta_bar_vmin})). Notice that, in order to do so, along
with $m_{\chi}$ and $\delta$ also the scaling law (either given by
Eq.(\ref{eq:scaling_law_f0}) or Eq. (\ref{eq:scaling_law_nlo})) is
required. According to the discussion above, in the
Isospin--Conserving case (i.e. for the scaling law of
Eq.(\ref{eq:scaling_law})) the upper bounds
$\bar{\tilde{\eta}}_{i,lim}^{SuperCDMS}$ from SuperCDMS in the same
$v_{min}$ ranges are well below the $\bar{\tilde{\eta}}_i^{CDMS-Si}$
estimations, so that, at face value and barring other issues such as
experimental systematic errors, an explanation of the CDMS--Si effect
in terms of WIMP scatterings is in strong tension with the available
data. However, as discussed in Section \ref{sec:isospin_violation}, an
appropriate choice of the parameters
$\bar{\lambda}_q,\bar{\lambda}_{\theta}$ can suppress the expected
rate on Germanium compared to that on Silicon, driving the
$\bar{\tilde{\eta}}_{i,lim}^{SuperCDMS}$ constraints above the
$\bar{\tilde{\eta}}_i^{CDMS-Si}$ estimations. Quantitatively, the
compatibility between the two results can be assessed introducing the
following compatibility ratio:

\begin{equation}
  {\cal D}(m_{\chi},\delta,\bar{\lambda}_d,\bar{\lambda}_s,\bar{\lambda}_{\theta}) \equiv \max_{i\in \mbox{signal}}
\left (\frac{\bar{\tilde{\eta}}_i^{CDMS-Si}+\sigma_i}{\min_{j\le i}\bar{\tilde{\eta}}_{j,lim}} \right ),
\label{eq:degrading_factor_eta_i}
\end{equation}

\noindent where $\sigma_i$ represents the standard deviation on
$\bar{\tilde{\eta}}_i^{CDMS-Si}$ as estimated from the data, $i\in
\mbox{signal}$ means that the maximum of the ratio in parenthesis is
for $v_{min,i}$ corresponding to the CDMS--$Si$ excess, while, due to
the fact that the function $\tilde{\eta}$ is non--decreasing in all
velocity bins $v_{min,i}$, the denominator contains the most
constraining bound on $\tilde{\eta}$ for $v_{min,j}\le v_{min,i}$. The
latter minimum includes all available bounds, although, in practice,
only Super--CDMS will prove to be effective in the discussion of
Section \ref{sec:results}.  Specifically, compatibility between
CDMS--Si and all other experiments (including SuperCDMS) is ensured if
${\cal D}<$1: in this case all the 1--$\sigma$ ranges of the
$\bar{\tilde{\eta}}_i^{CDMS-Si}$'s are below the upper bounds from
other experiments with $v_{min}\le v_{min,i}$. Notice that the
definition above combines different $v_{min,i}$ bins, allowing for
some energy--dependence in the scaling law (as suggested by
Eqs.(\ref{eq:diff_rate_nlo}) and (\ref{eq:nlo_amplitudes})). On the
other hand, if only one target is relevant for the constraint and in
the LO case, or if the energy dependence is neglected in the NLO case,
the ratio between the scaling laws corresponding to Silicon and the
target element of the bound factors out in the sum above. In this case
${\cal D}$ is a function of
$\bar{\lambda}_d$,$\bar{\lambda}_s$,$\bar{\lambda}_{\theta}$ only
through particular combinations (the $r=f_n/f_p$ parameter in the LO
case or the $t$ parameter defined in
Eq.(\ref{eq:scaling_law_t})). However, if the scaling law depends on
the energy, all the three ratios
$\bar{\lambda}_d$,$\bar{\lambda}_s$,$\bar{\lambda}_{\theta}$ are
needed to calculate ${\cal D}$. In the former case ${\cal D}$ can be
minimized as a function of $r$ or $t$ and its minimum value ${\cal
  D}_{min}$ is constant in the plane
$\bar{\lambda}_s$--$\bar{\lambda}_{\theta}$ (although each point will
correspond to a different value of $\bar{\lambda}_{d,min}$). In the
latter case ${\cal
  D}(\bar{\lambda}_d,\bar{\lambda}_s,\bar{\lambda}_{\theta})$ can be
minimized as a function of $\bar{\lambda}_d$ at fixed
$\bar{\lambda}_s$ and $\bar{\lambda}_{\theta}$.  We will proceed in
this way in the numerical analysis of Section \ref{sec:results}.

We also notice here that the $\bar{\tilde{\eta}}_i^{CDMS-Si}$ values
determined from the data must be compatible with the property that the
$\tilde{\eta}$ function is decreasing with $v_{min}$. A possible
criterion to quantify this condition is that the lower range of each
$\bar{\tilde{\eta}}_{i+1}^{CDMS-Si}$ falls below the upper range of
the $\bar{\tilde{\eta}}_{i}^{CDMS-Si}$ with immediately lower
$v_{min}$, i.e.:

\begin{equation} {\cal R}=\max_{i\in
    \mbox{signal}-1}\frac{\bar{\tilde{\eta}}_{i+1}^{CDMS-Si}-\sigma_{i+1}}{\bar{\tilde{\eta}}_i^{CDMS-Si}+\sigma_i}<1.
\label{eq:decreasing}
\end{equation}

\noindent Combining conditions (\ref{eq:degrading_factor_eta_i}) and
(\ref{eq:decreasing}), full compatibility is achieved if $\max({\cal
  D},{\cal R})<1$. In practice, due to the low statistics of the
CDMS--Si excess and the ensuing large error bars on the
$\bar{\tilde{\eta}}_i^{CDMS-Si}$, ${\cal R}$ is always less than 1 and
only the condition (\ref{eq:degrading_factor_eta_i}) will turn out to
be effective (see for instance Fig.\ref{fig:eta_vmin}).

Once the $m_{\chi}$, $\delta$, $\bar{\lambda}_d$, $\bar{\lambda}_s$,
$\bar{\lambda}_{\theta}$ parameters are fixed, the value of the
suppression scale $\tilde{\Lambda}$ required to explain the
$\bar{\tilde{\eta}}_i^{CDMS-Si}$ obtained from the CDMS--Si data
should be estimated in order to get information on the underlying DM
model. This requires to determine the cross section $\sigma_0$ from
Eq.(\ref{eq:eta_tilde}), disentangling particle physics form
astrophysics, and is not possible without specifying the velocity
distribution $f(\vec{v})$. However, since $\int f(\vec{v}) d^3 v$=1,
the function $\eta(v_{min})$ can be at least maximized by the choice
$f(\vec{v})=\delta(v_s-v_{min})$, with $v_s$ the maximal value of
the $v_{min}$ range corresponding to the CDMS--Si excess. This
corresponds to the largest $\eta(v_{min})$ which does not vanish in
the signal range. In this case:

\begin{equation}
\tilde{\eta}^{max}(v_{min})=\bar{\tilde{\eta}}_{fit}^{CDMS-Si} \theta(v_s-v_{min}).
\label{eq:eta_max_vs}
\end{equation}

\noindent 
Notice that the large error bars on the
$\bar{\tilde{\eta}}_{i}^{CDMS-Si}$ imply that, indeed, the above
ensuing flat functional form for the $\tilde{\eta}$ function is not
incompatible with the data.  The constant value
$\bar{\tilde{\eta}}_{fit}^{CDMS-Si}$ can then the be fitted from the
data in a straightforward way. Plugging Eq.(\ref{eq:eta_max_vs}) in
Eq.(\ref{eq:eta_tilde}) and using for $\sigma_0$ the expression given
in Eq.(\ref{eq:sigma_0}), one gets:

\begin{equation}
  \tilde{\Lambda}=f_0^{\frac{1}{3}}\left(\frac{2 \rho_{\chi} \mu_{\chi p}^2}{\pi \bar{\tilde{\eta}}_{fit}^{CDMS-Si} m_{\chi} v_s} \right )^{\frac{1}{6}}.
\label{eq:lambda_tilde}
\end{equation}

Obviously $\tilde{\Lambda}$ does not depend on choice of the arbitrary
amplitude $f_0$ chosen for the factorization, since
$\bar{\tilde{\eta}}_{fit}^{CDMS-Si}$ is fitted using the scaling law
(\ref{eq:scaling_law_f0}) and scales with $f_0^2$.

Notice than in Eq.(\ref{eq:eff_lagrangian_full}) for dimensional
reason the suppression scale $\Lambda$ appears at the third power
having extracted a quark mass factor from the coupling. We followed
this convention to comply to that of
Ref. \cite{isospin_violation_nlo2}. An alternative way to parametrize
the same interaction is to write the Lagrangian in the form:

\begin{equation}
  {\cal L}=\sum_{q=u,d,s,c,b,t} \frac{\tilde{\xi}_q}{M_*^2}\bar{\chi}^{\prime}\chi \bar{q} q \mbox{+h.c.}.
\label{eq:eff_lagrangian_m_star}
\end{equation}

\noindent Lacking a knowledge of the ultraviolet completion of the
model both forms are acceptable. However, since $m_q/\Lambda\ll$1,
numerically $M_*\gg \Lambda$ for the same values of the expected
signals, so in order to get an upper bound on the scale of the new
physics involved in the process we chose to discuss constraints on
$M_*$. In particular this can be done by imposing the perturbativity
condition:

\begin{equation}
  \max(|\tilde{\xi}_q|)\le4 \pi,
\label{eq:pertubativity}
\end{equation}

\noindent which implies:

\begin{equation}
  M_*\le \sqrt{4\pi} \tilde{\Lambda} \left (\frac{\tilde{\Lambda}}{\mbox{max($m_q \bar{\lambda}_q$)}} \right )^{\frac{1}{2}},
\label{eq:m_star_max}
\end{equation}

\noindent with $\tilde{\Lambda}$ given by Eq.(\ref{eq:lambda_tilde}).

The procedure outlined above will be adopted In Section
\ref{sec:results} to get an estimation of the maximal value of
$M_*$. 

\section{Relic abundance}
\label{sec:relic_abundance}

A minimal necessary requirement of the exothermic DM scenario
(i.e. $\delta<0$) is that the metastable $\chi$ particle decays to the
lower--mass state $\chi^{\prime}$ on a time scale larger than the age
of the Universe. Specifically, the effective Lagrangian of
Eq.(\ref{eq:eff_lagrangian_full}) drives the decay $\chi\rightarrow
\chi^{\prime} \gamma\gamma$, whose amplitude has been recently
estimated making use of Chiral Perturbation Theory \cite{decay}:

\begin{equation}
\Gamma\simeq \frac{B_0^2 \alpha_{EM}^2}{32 (105 \pi^5) \Lambda_c^4 \tilde{\Lambda}^6}
\left| m_u\tilde{\lambda}_u-m_d\tilde{\lambda}_d   \right|^2 \delta^7,
\label{eq:decay}
\end{equation}

\noindent with $B_0=m_{\pi}^2/(m_u+m_d)$, $\Lambda_c\simeq 4\pi
f_{\pi}/\sqrt{N}$, $N=2$, $f_{\pi}$=93 MeV and $m_{\pi}$ the pion
mass. As discussed in \cite{decay}, the range of the $\delta$
parameter involved in direct detection ($|\delta|\lsim$ 100 keV)
drives the amplitude (\ref{eq:decay}) to values safely below that
corresponding to the age of the Universe, $\tau_U^{-1}\simeq 1.5\times
10^{-42}$ GeV \footnote{Also indirect signals from present $\chi$
  decays detectable in the hard $X$--ray spectrum of the galactic
  diffuse gamma background are strongly suppressed in this
  regime\cite{decay}.}. This will be confirmed by the quantitative analysis of Section \ref{sec:results}.

As far as the production mechanism for the $\chi$ particle
cosmological density is concerned, several can be devised. Thermal
decoupling, where the particles $\chi$ and $\chi^{\prime}$ are
initially in thermal equilibrium in the plasma of the Early Universe
until their interactions freeze-out at a temperature $T\simeq
m_{\chi}/20$, is the most standard and predictive. One can notice that
the mass splitting involved ($|\delta|\lsim$ 100 keV) is much smaller
than the typical freeze--out temperature even for very light WIMPs,
$T_f\simeq m_{\chi}/20\gsim$ 50 MeV for $m_{\chi}\gsim$ 1 GeV. As a
consequence of this the chemical potential between the $\chi$ and
$\chi^{\prime}$ states can be safely neglected, and the corresponding
number (and mass) densities $n_{\chi}$ and $n_{\chi^{\prime}}$ of both
species should be the same when they decouple at $T=T_F$.  This is
also true below $T_F$ when, as long as $T\gsim \delta$, kinetic
equilibrium is maintained by the reactions
$\chi+q\leftrightarrow\chi^{\prime}+q$ (below the QCD--phase
transition, $T_{QCD}\simeq 150$ MeV, these reactions will briefly
involve pions until the latter become non--relativistic and their
density drops exponentially). The bottom line is that, being
$\chi^{\prime}$ stable on cosmological scales, the condition
$n_{\chi}$=$n_{\chi^{\prime}}$ is likely to be maintained until the
present day, so that, in this scenario, galactic halos contain equal
densities of $\chi$ and $\chi^{\prime}$. In principle this implies
that direct detection experiments should be able to measure {\it at
  the same time} $\chi$ down--scatters to $\chi^{\prime}$ and
$\chi^{\prime}$ up--scatters to $\chi$. However, we notice that, as
discussed in Section \ref{sec:cdms-si}, if exothermic DM is advocated
to explain the three WIMP candidates in CDMS--$Si$, very low WIMP
masses are required, $m_{\chi}\lsim$ 4 GeV, as well as $\delta\lsim
-50$ keV \cite{noi}. In this case Eq. (\ref{eq:vmin} ) implies that
up-scatters ($\delta>0$) are only possible when the incoming WIMP
velocity (in the Earth's rest frame) is larger than the minimal value
$v_{min}=v_{*}=\sqrt{2 \delta/\mu_{\chi N}}\simeq \sqrt{2
  \delta/m_{\chi}}\gsim$ 950 km/s, a value incompatible to
acceptable values of the galactic escape velocity (in Section
\ref{sec:cdms-si} we adopt the reference value $v_{esc}^{Lab}\simeq$
782 km/s). So, in this particular scenario, only down--scatters are
detectable, and this implies that strictly speaking, if
$\rho_{DM}=\rho_{\chi}+\rho_{\chi^{\prime}}$ is the DM density
estimated in the neighborhood of the Sun by its gravitational
interactions, the incoming WIMP flux to which {\it all direct
  detection experiments} are sensitive is proportional to only half of
that, $\rho_{DM}/2$.

As in the elastic case\cite{cdms_si_eff,cdms_si_eff2}, the coannihilation cross
section for the process $\chi\chi^{\prime}\rightarrow q q$ at
decoupling is velocity--suppressed for a scalar interaction because in
$s$--wave the $\chi\chi^{\prime}$ system has parity -1.
As a consequence, if the typical scale $\tilde{\Lambda}$ required to
explain the CDMS--Si excess is used to calculate the relic abundance,
too large values incompatible to observation are found
\cite{cdms_si_eff}. This will be generally confirmed by the
phenomenological analysis of Section \ref{sec:results}. However, as
shown in the previous Section, in the isospin--violating scenario the
regions of the ($\bar{\lambda}_{\theta}$--$\bar{\lambda}_s$) parameter
space close to Eq.(\ref{eq:straight_line}) or
Eq. (\ref{eq:straight_line_nlo}) correspond to large cancellations
among the coupling constants $\tilde{\lambda}_q$ in the calculation of
the direct detection rate, which drive the suppression scale
$\tilde{\Lambda}$ to values significantly lower than in the
isospin--conserving case for a fixed number of predicted events. On
the other hand, in the calculation of the coannihilation cross section
that fixes the thermal relic abundance only the absolute values of the
coupling constants $|\tilde{\lambda}_q|^2$ appear:

\begin{eqnarray}
<\sigma v>&=&\frac{3 m_{\chi}^2}{8\pi \Lambda^6}
\sum_q m_q^2 \left |\tilde{\lambda}_q \right |^2 
\left ( 1-\frac{m_q^2}{m_{\chi}^2}  \right )^{\frac{3}{2}} \times \nonumber\\ 
&&\left (6 x^{-1} -27 x^{-2}+... \right ),
\label{eq:sigmav}
\end{eqnarray}

\noindent so that no cancellation is at work to compensate the
enhancement due to a small $\tilde{\Lambda}$. As we will see, this
represents a possible mechanism to drive the thermal relic abundance
down to values compatible to observation also for a scalar--type
interaction, at variance with the Isospin--conserving case. In
Eq.(\ref{eq:sigmav}) $x\equiv m_{\chi}/T$, and we have neglected the
mass splitting between $\chi$ and $\chi^{\prime}$ taking
$m_{\chi}^{\prime}$=$m_{\chi}$ since $\delta \ll T_f\simeq
m_{\chi}/20$. Below $T_f$ the total relic density of $\chi$ and
$\chi^{\prime}$ freezes to the usual value:

\begin{equation}
\Omega h^2 \simeq \frac{8.7\times 10^{-11}/\mbox{GeV$^2$}} {g_*^{1/2}(x_f)} \frac{1}{\int_0^{x_f} <\sigma v> dx},
\label{eq:omega}
\end{equation}

\noindent where $g_*(x_f)$ denotes the number of relativistic degrees
of freedom of the thermodynamic bath at $x_f$.

\section{Signals at the LHC}
\label{sec:LHC}

We have simulated monojet+missing transverse energy $\not\!E_T$
\cite{monojet_theory,monojet_exp} and hadronically--decaying mono
$W/Z$ events\cite{mono_wz_theory,atlas_wz} at the LHC in $pp$
collisions at $\sqrt{s}$ = 8 TeV using MadGraph 5 \cite{madgraph},
interfaced with Pythia 6 \cite{pythia} and Delphes 3 \cite{delphes},
with the scalar interaction of Eq.(\ref{eq:eff_lagrangian_full}),
using CTEQ6L1 Parton Distribution Functions \cite{cteq6l1} and
including the $b$ quark.  Notice that in collider physics the effect
of the mass splitting between $\chi$ and $\chi^{\prime}$ is completely
negligible.

As far as the monojet signal is concerned, we have required the
transverse momentum $P_t$ of each parton to be larger than 80 GeV
\cite{pt_jet} and applied the following kinematic cuts at the detector
level \cite{cdms_si_eff2}:

\begin{itemize}

\item $p_{T_{j}} >$ 110 GeV, with $p_{T_j}$ the jet transverse
  momentum;

\item $|\eta_j|<$  2.4, with $|\eta_j|$ the jet pseudorapidity; 

\item $\not\!E_T >$ 400 GeV.

\end{itemize}

In the case of hadronically--decaying mono $W/Z$ events, we calculate
the total production cross section, applying the same cuts used in \cite{atlas_wz}:

\begin{itemize}

\item $p^{W,Z} _T >$ 250 GeV, where $p^{W,Z}_T$ is the $W$ or $Z$ transverse momentum;

\item $|\eta|^{W,Z} <$ 1.2, where $\eta^{W,Z}$ is the pseudo-rapidity;

\item $\sqrt{y} >$ 0.4, where $\sqrt{y} \equiv \min(p^T_1, p^T_2)
  \Delta R/m_{jet}$, with $p^T_i$ (i = 1 or 2) being the transverse
  momentum of the two leading reconstructed jets from the $W$ or $Z$
  decay, $\Delta R$=$\sqrt{(\Delta \eta)^2+(\Delta \phi)^2}$ the
  distance in pseudo--rapidity and azimuthal angle between jets, and
  $m_{jet}$ the jets invariant mass.

\item $P^T_{\chi \chi^{\prime}} >$ 350 GeV, with $P^T_{\chi
    \chi^{\prime}}$ the invisible transverse momentum carried away by
  the $\chi$, $\chi^{\prime}$ particles.

\end{itemize}

In both analyzes we have used the anti--kt jet--reconstruction algorithm
with parameter $R$=0.5.

\section{Discussion}
\label{sec:results}

\begin{figure*}[ht]
\begin{center}
\includegraphics[width=1\columnwidth,bb= 46 194 506 635]{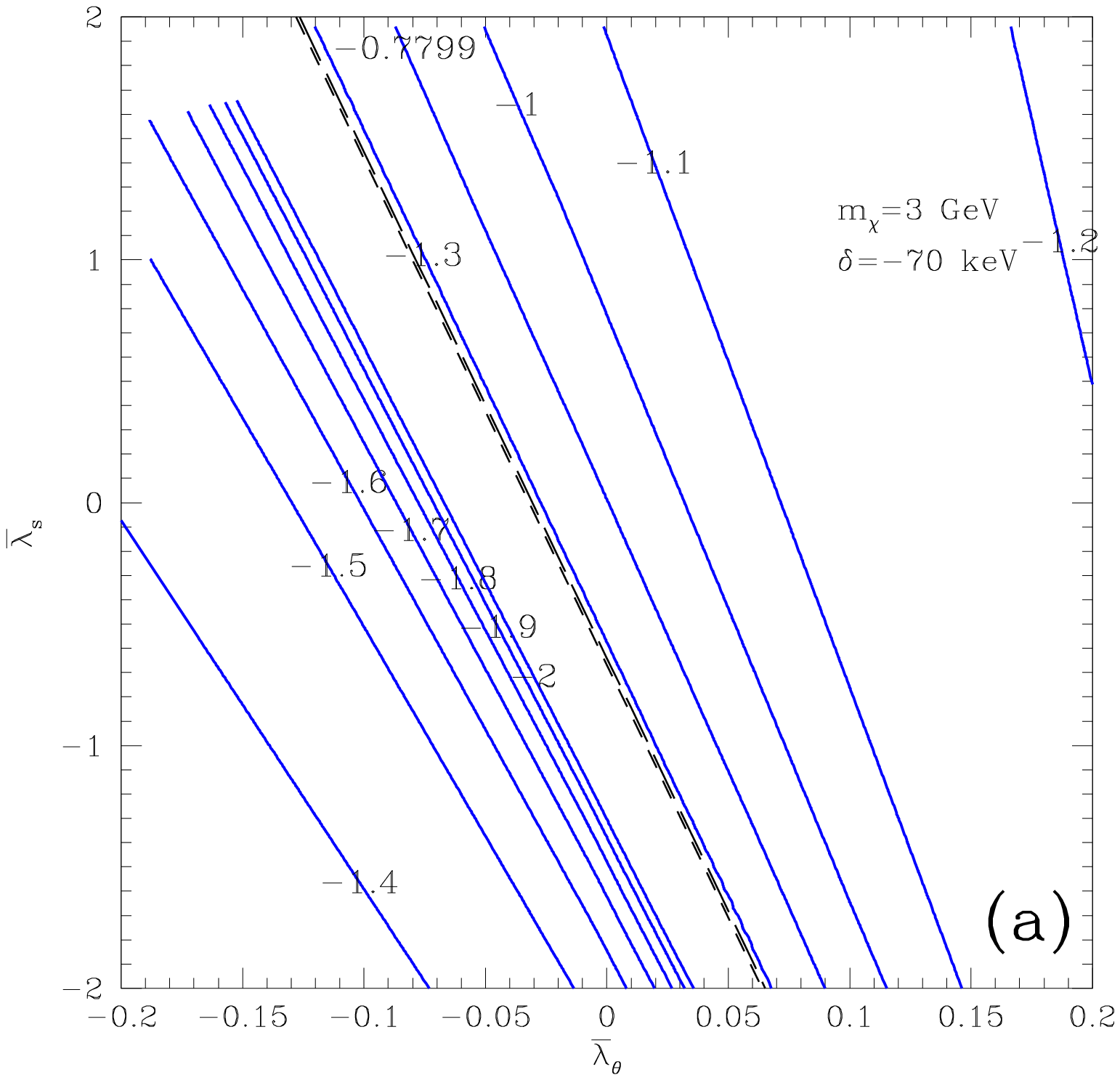}
\includegraphics[width=1\columnwidth,bb= 46 194 506 635]{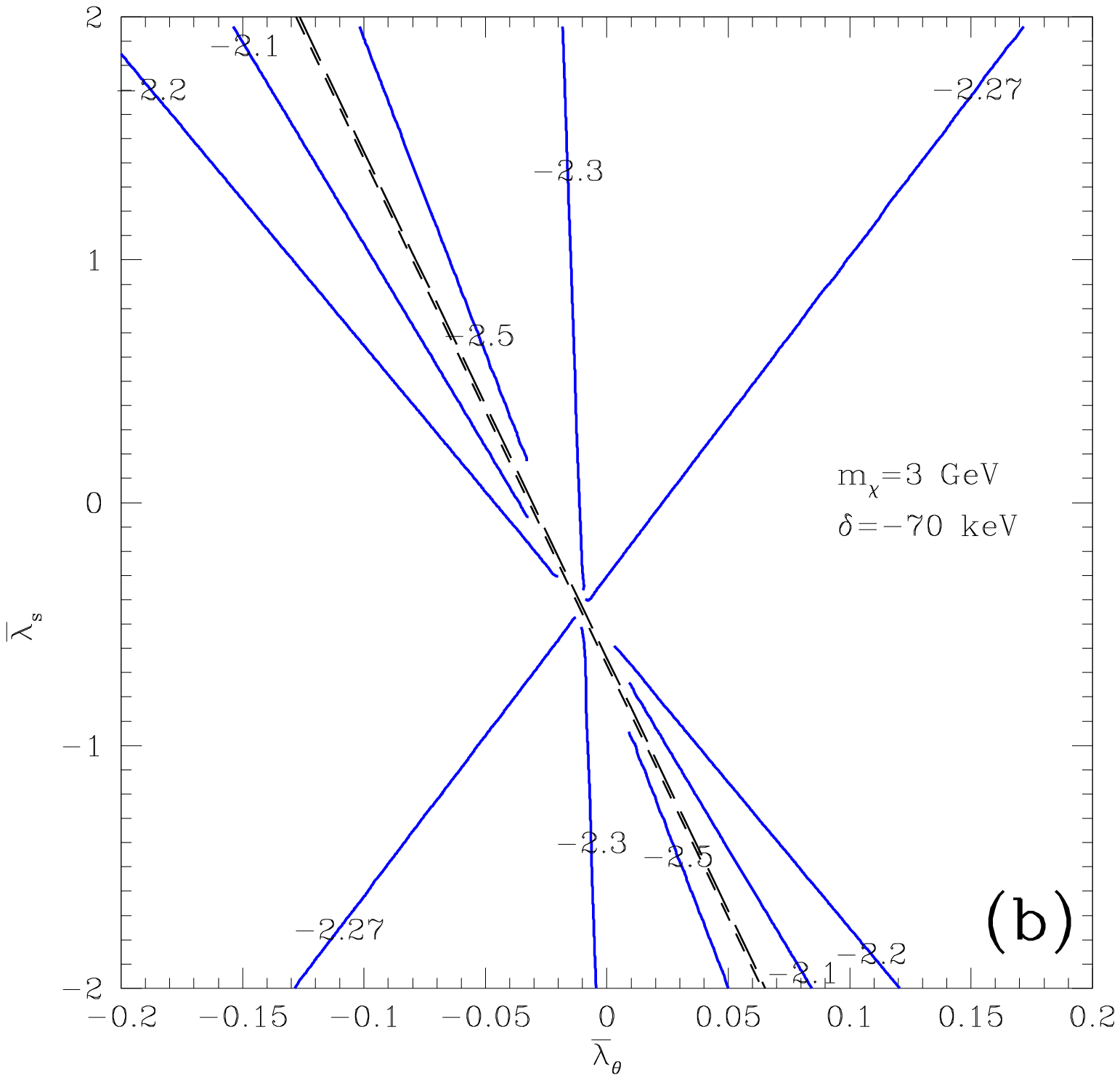}
\end{center}
\caption{Isospin--violation parameter space for $m_{\chi}$=3 GeV and
  $\delta$=-70 keV (IDM benchmark point indicated with a (green) cross
  in Fig. \ref{fig:mchi_delta_si}). For each value of
  $\bar{\lambda}_{\theta}$ and $\bar{\lambda}_{s}$ the remaining
  parameter $\bar{\lambda}_{d}$ is set to the value
  $\bar{\lambda}_{d,min}$ minimizing the quantity ${\cal D}$ defined
  in Eq. (\ref{eq:degrading_factor_eta_i}). {\bf (a)} The solid lines
  show constant values of $r=f_n/f_p=r_{min}$, i.e. the value of $r$
  that minimizes the compatibility ratio ${\cal D}$. {\bf (b)} The
  solid lines show constant values of $t=t_{min}$, i.e. the value of
  the $t$ parameter introduced in Eq. (\ref{eq:scaling_law_t}) that minimizes the
  compatibility ratio ${\cal D}$. In both figures the NLO corrections
  of Eq. (\ref{eq:diff_rate_nlo}) are included, while the
  short--dashed and long dashed straight lines represent the
  ``alignment'' conditions given in Eqs.(\ref{eq:straight_line}) and
  (\ref{eq:straight_line_nlo}), respectively.}
\label{fig:r_fn_fp}
\end{figure*}

\begin{figure}[ht]
\begin{center}
\includegraphics[width=1\columnwidth,bb= 46 194 506 635]{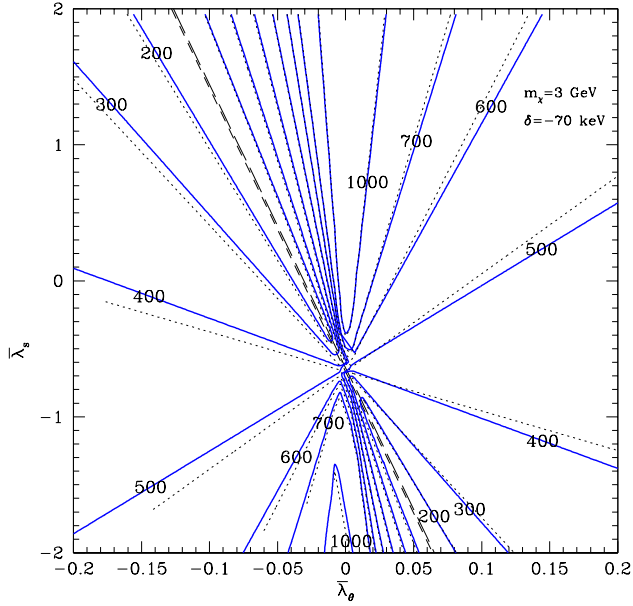}
\end{center}
\caption{Same as in Fig. \ref{fig:r_fn_fp}. On the solid lines the
  upper bound on the suppression scale $M_*$ (introduced in
  Eq.(\ref{eq:m_star_max})) is fixed to the indicated value when the NLO
  corrections of Eq. (\ref{eq:diff_rate_nlo}) to the expected WIMP
  rate are included. Dashed lines represent the same for the LO
  calculation.}
\label{fig:lambda_theta_m_star}
\end{figure}

\begin{figure}[ht]
\begin{center}
\includegraphics[width=1\columnwidth,bb= 46 194 506 635]{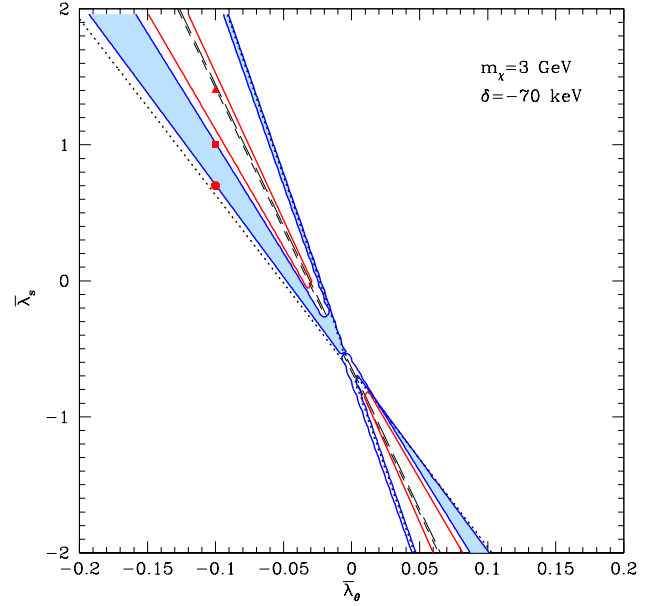}
\end{center}
\caption{Same as in Fig.\ref{fig:r_fn_fp}.  The shaded regions are
  bounded by the solid blue lines corresponding to the two conditions
  $\Omega_{\chi}h^2$=0.12 and ${\cal D}_{min}$=1 and represent the
  parameter space where CDMS--Si and SuperCDMS are mutually compatible
  while at the same time the metastable state $\chi$ can be a thermal
  relic (i.e. $\Omega_{\chi}h^2\le$0.12 using
  Eq. (\ref{eq:omega})). The dotted curve represents the condition
  $\Omega_{\chi}h^2$=0.12 calculated using the LO scaling law for the
  expected rate (see Eq.(\ref{eq:diff_rate}). Finally, the inner solid
  (red) line corresponds to $\tau=1/\Gamma$=4$\times$10$^{26}$
  seconds, as given by Eq.(\ref{eq:decay}). The (red) circle indicates
  the representative choice of $\bar{\lambda}_{\theta}$,
  $\bar{\lambda}_{s}$ shown in Fig.\ref{fig:eta_vmin}(a) for which
  $\Omega_{\chi}h^2$=0.12, which corresponds to ${\cal D}_{min}$=0.7;
  the square indicates the choice of $\bar{\lambda}_{\theta}$,
  $\bar{\lambda}_{s}$ shown in Figs.\ref{fig:eta_vmin}(b)
  \ref{fig:degrading_factor}(a) and for which ${\cal D}_{min}$=1
  (i.e. at the verge of incompatibility between the CDMS--Si result
  and the SuperCDMS constraint); the (red) triangle indicates the
  representative choice of $\bar{\lambda}_{\theta}$,
  $\bar{\lambda}_{s}$ shown in Figs.\ref{fig:degrading_factor}(b), and
  exemplifies a configuration very close to
  Eq.(\ref{eq:straight_line_nlo}) where the NLO energy--dependent
  corrections of Eq. (\ref{eq:diff_rate_nlo}) spoil the maximal
  achievable cancellation in WIMP--$Ge$ scattering.}
\label{fig:allowed_region}
\end{figure}

\begin{figure*}[ht]
\begin{center}
\includegraphics[width=1\columnwidth,bb= 46 194 506 635]{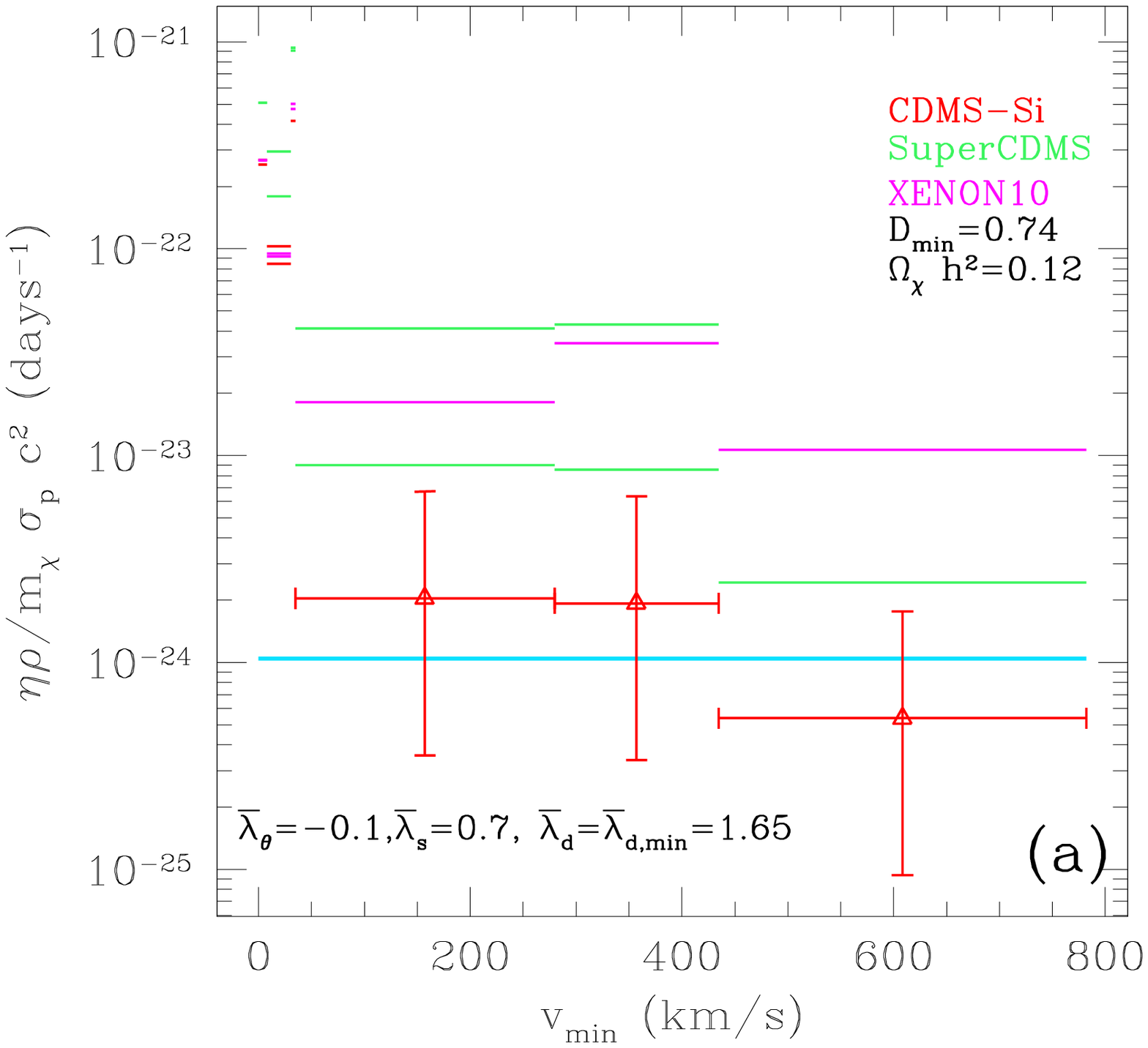}
\includegraphics[width=1\columnwidth,bb= 46 194 506 635]{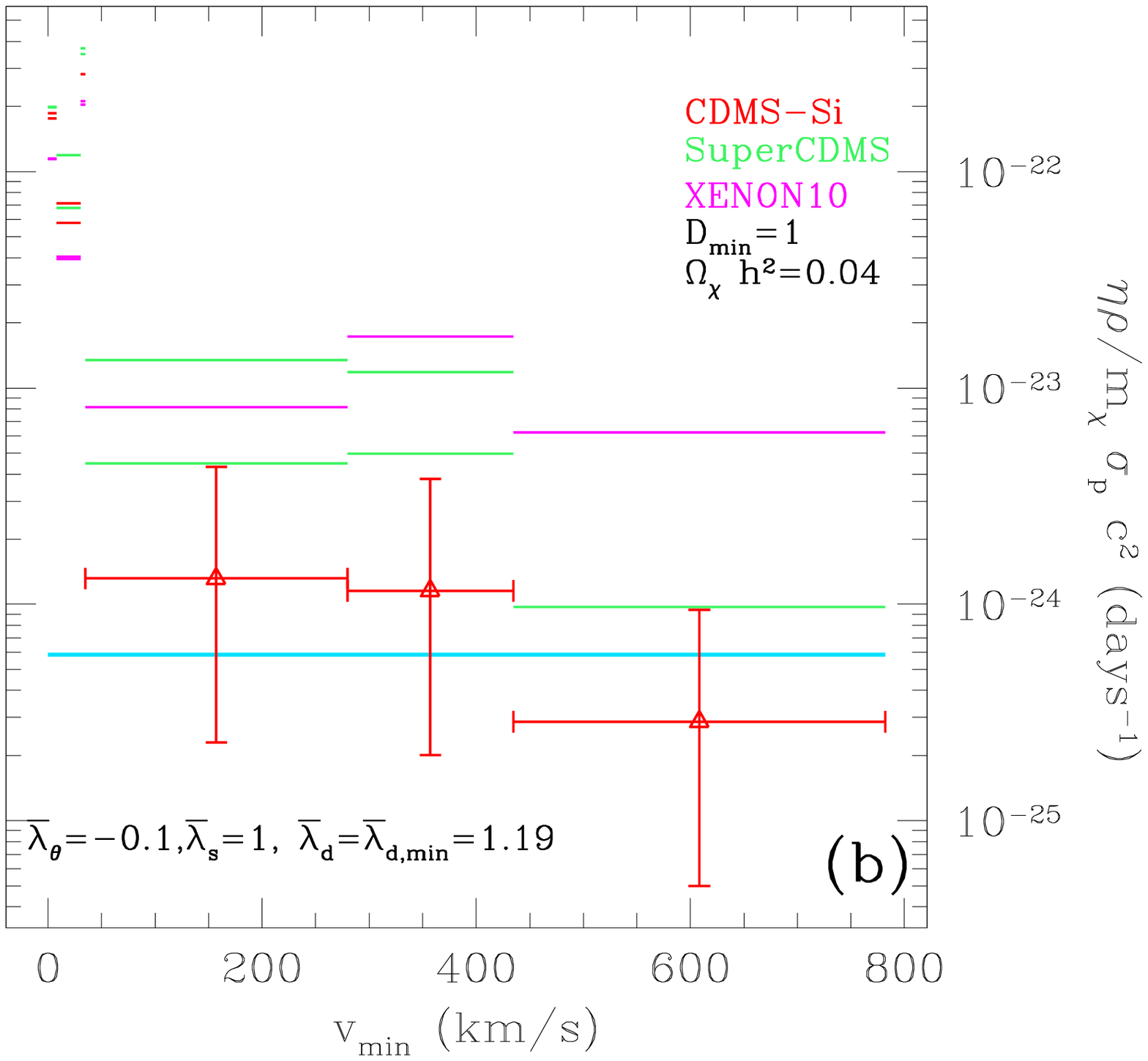}
\end{center}
\caption{Measurements and bounds for the function $\tilde{\eta}$
  defined in Eq.(\protect\ref{eq:eta_tilde}) for $\sigma_0=\sigma_p$
  and $m_{DM}$=3 GeV, $\delta$=-70 keV, i.e. for the IDM benchmark
  point indicated with a (green) cross in
  Fig. \ref{fig:mchi_delta_si}. {\bf (a)}
  $\bar{\lambda}_{\theta}$=-0.1, $\bar{\lambda}_{s}$=0.7 (benchmark
  point indicated with a circle in Fig.\ref{fig:allowed_region},
  corresponding to $\Omega_{\chi}h^2$=0.12). {\bf (b)}
  $\bar{\lambda}_{\theta}$=-0.1, $\bar{\lambda}_{s}$=1 (benchmark
  point indicated with a square in Fig.\ref{fig:allowed_region} and
  corresponding to ${\cal D}_{min}$=1). In both figures
  $\bar{\lambda}_{d}$=$\bar{\lambda}_{d,min}$ minimizes the ${\cal D}$
  function of Eq. (\ref{eq:degrading_factor_eta_i}) and the thick
  (blue) horizontal line represents the quantity
  $\bar{\tilde{\eta}}_{fit}^{CDMS-Si}$ introduced in
  Eq. (\ref{eq:eta_max_vs}). Note that, consistently with the
  condition ${\cal D}_{min}$=1, in plot (b) the constraint from
  SuperCDMS ``touches'' the upper range of the CDMS--$Si$ excess.}
\label{fig:eta_vmin}
\end{figure*}

\begin{figure*}[ht]
\begin{center}
\includegraphics[width=1\columnwidth,bb= 46 194 506 635]{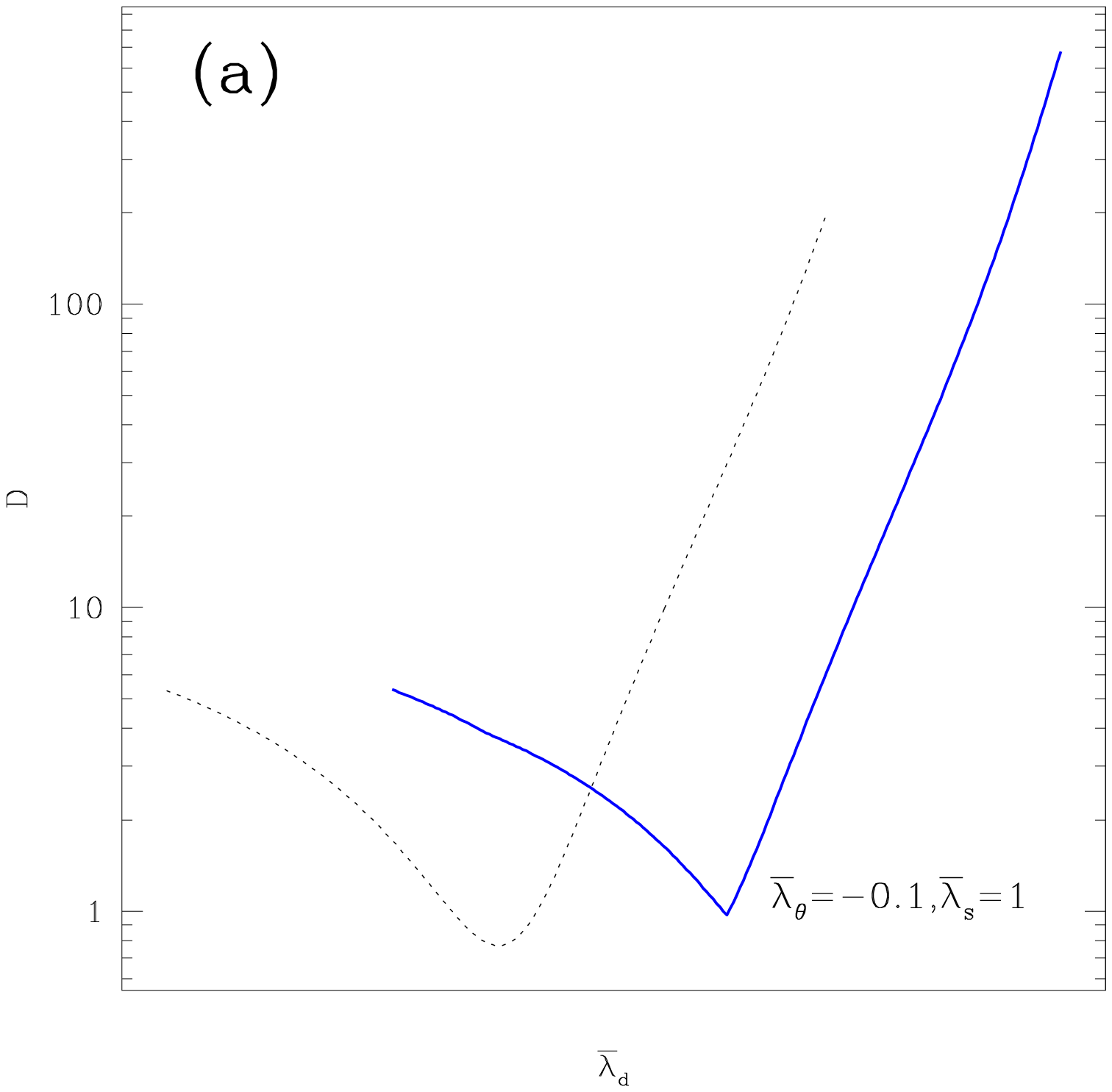}
\includegraphics[width=1\columnwidth,bb= 46 194 506 635]{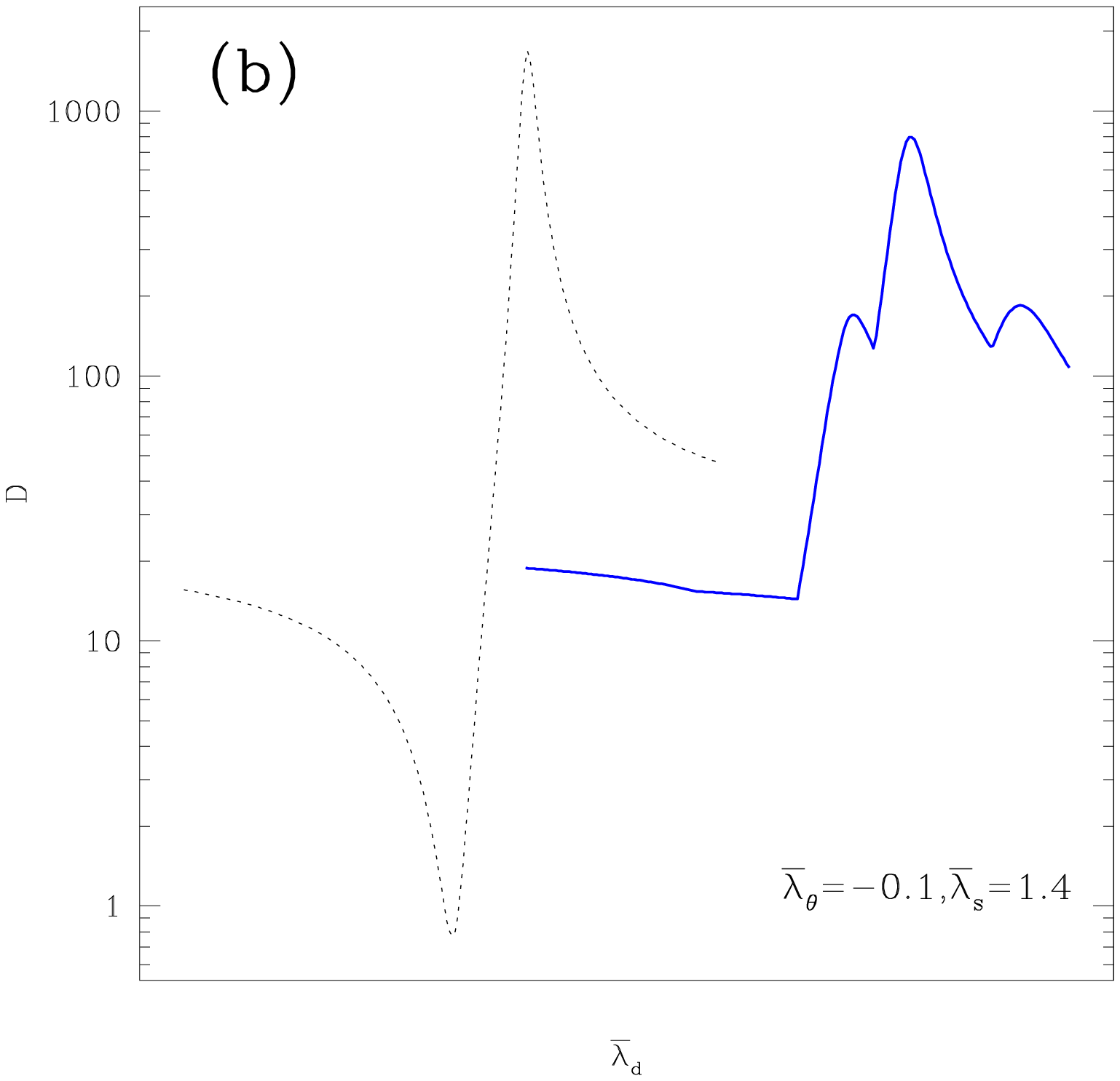}
\end{center}
\caption{The compatibility ratio ${\cal D}$ defined in
  Eq.(\ref{eq:degrading_factor_eta_i}) is plotted as a function of
  $\bar{\lambda}_d$ for $m_{DM}$=3 GeV, $\delta$=-70 keV, i.e. for
  the IDM benchmark point indicated with a (green) cross in
  Fig. \ref{fig:mchi_delta_si}.{\bf (a)}
  $\bar{\lambda}_{\theta}$=-0.1, $\bar{\lambda}_s$=1 (benchmark point
  indicated with a (red) square in
  Fig.(\ref{fig:allowed_region})). {\bf (b)}
  $\bar{\lambda}_{\theta}$=-0.1, $\bar{\lambda}_s$=1.4 (benchmark
  point indicated with a (red) triangle in
  Fig.(\ref{fig:allowed_region})). In both figures the solid (blue)
  line represents the calculation including the NLO corrections of
  Eq. (\ref{eq:diff_rate_nlo}), while the thin dotted (black) line
  shows the same quantity when the approximate expression for the NLO
  corrections of Eq.(\ref{eq:diff_rate_nlo_approx}) is used, which
  neglects the terms with explicit energy dependence.}
\label{fig:degrading_factor}
\end{figure*}

\begin{figure}[ht]
\begin{center}
\includegraphics[width=1\columnwidth,bb= 46 194 506 635]{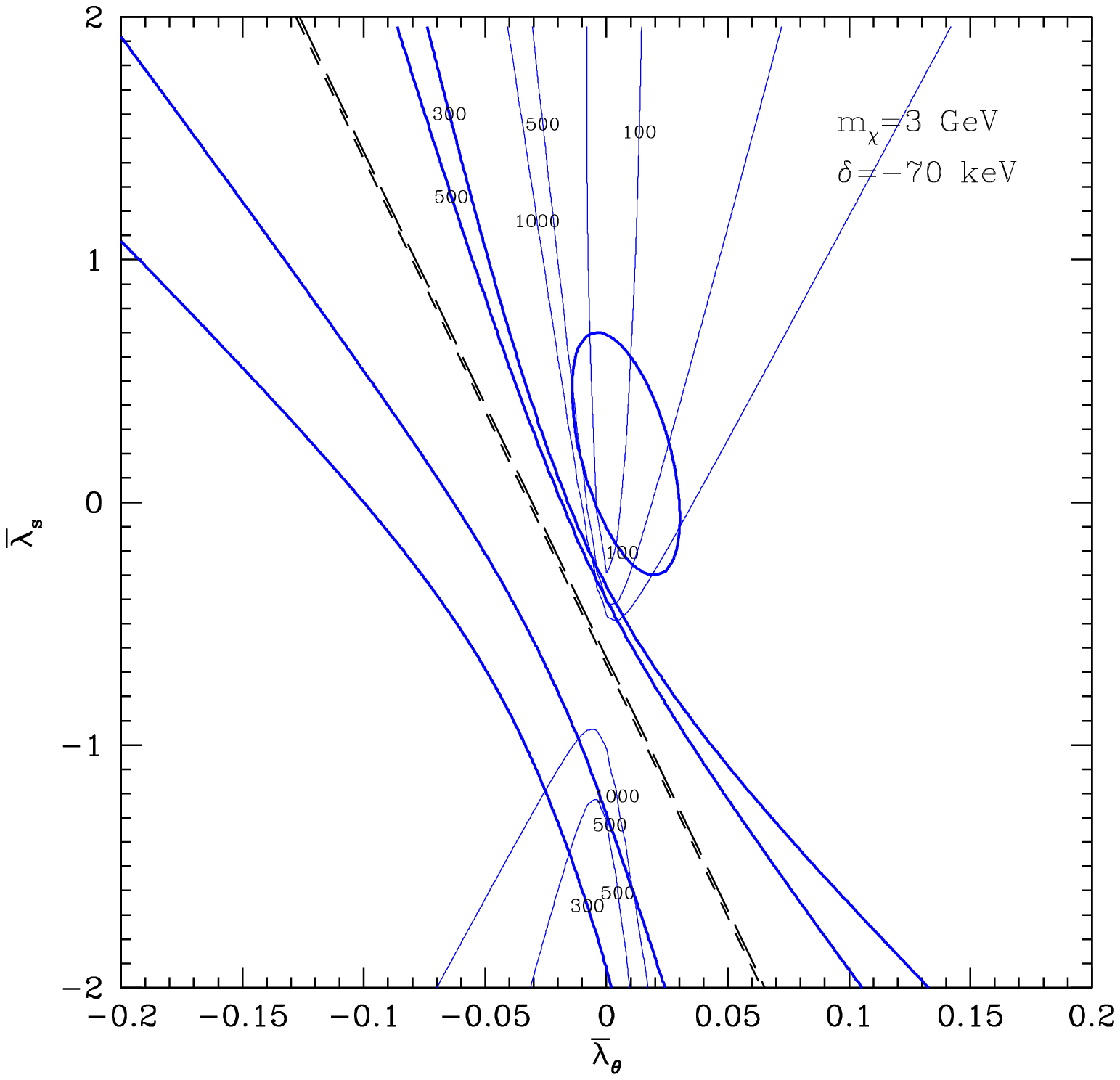}
\end{center}
\caption{Same as in Fig.\ref{fig:r_fn_fp}. The thick solid lines show
  constant values of the cross section for hadronically--decaying mono
  $W/Z$ events in $fb$, while thin solid lines represent constant
  values of the expected number of monojet+missing energy events for
  the integrated luminosity ${\cal L}$=19.5 $fb^{-1}$. In both cases
  $\sqrt{s}$=8 TeV. The applied kinematic cuts are listed in Section
  \ref{sec:LHC}.}
\label{fig:lhc}
\end{figure}

In the following we will assume that the CDMS--$Si$ effect is
explained in terms of a WIMP signal, and we will fix the IDM
parameters to the benchmark $m_{\chi}$=3 GeV, $\delta$=-70 keV
(indicated by a cross in Fig. \ref{fig:mchi_delta_si}). As already
explained, this choice of parameters is kinematically not accessible
to XENON100 and LUX, but is covered by SuperCDMS \cite{noi}, so that
the Isospin Violation mechanism must be advocated to suppress WIMP
scattering on $Ge$ targets compared to $Si$, or, quantitatively,
minimizing the compatibility ratio ${\cal D}$ introduced in
Eq.(\ref{eq:degrading_factor_eta_i}). In order to calculate the
latter, in our analysis we will fix the following three energy bins
for CDMS-$Si$, each containing one of the WIMP candidate events: 7.4
keVnr$\le$8.2 keVnr, 8.2 keVnr$\le$9.5 keVnr, 9.5 keVnr$\le$12.3
keVnr\footnote{For our particular choice of $m_{\chi}$ and $\delta$
  this binning ensures that the mapping between energies and $v_{min}$
  is univocal both for CDMS--$Si$ and for SuperCDMS.}. We will then
map them through $v_{min}$ into energy ranges for SuperCDMS, for which
we use the low--energy analysis of Ref.\cite{super_cdms} with a
Germanium target in the energy range 1.6 keVnr$<E_R<$ 10 keVnr with a
total exposition of 577 kg day and 11 observed WIMP candidates. Since
the energy resolution in CDMS-$Si$ has not been measured, we take for
both CDMS-$Si$ and SuperCDMS
$\sigma_{CDMS-Si}(E^{\prime})=\sqrt{0.293^2+0.056^2
  (E^{\prime}/\mbox{keVnr})}$ in keVnr from \cite{cdms_resolution}.
As already pointed out, the definition of ${\cal D}$ includes in
principle all available bounds (also for the $v_{min}$ range {\it
  below} that corresponding to the CDMS--$Si$ effect). Specifically
the kinematic regime we are interested in is also probed by
XENON10. The latter makes use of the secondary ionization signal $S_2$
only, with an exposition of 12.5 day and a fiducial mass of 1.2 kg. In
the following we have included XENON10 in our numerical calculation of
${\cal D}$, although we have found that only Super--CDMS is relevant
to the discussion, since XENON10 does not imply any constraint (see
for instance Fig.\ref{fig:eta_vmin}).  In particular, for XENON10 we
have taken the scale of the recoil energy $E_R$ and the recorded event
spectrum in the energy range 1.4 keVnr$<E_R<$ 10 keVnr directly from
Fig. 2 of Ref. \cite{xenon10}, while for the energy resolution we have
assumed $\sigma_{XENON10}=E_R/\sqrt{E_R Q_y(E_R)}$ where $Q_y(E_R)$ is
the electron yield that we calculated with the same choice of
parameters as in Fig. 1 of \cite{xenon10}.

When the LO expression (\ref{eq:diff_rate}) is used for the WIMP
expected rate, ${\cal D}$ is minimized by
$r=f_n/f_p\simeq-0.78$\cite{isospin_violation}.  This situation is
modified when NLO corrections to the scattering rate (see
Eq.(\ref{eq:diff_rate_nlo})) are included. This is shown in Fig
\ref{fig:r_fn_fp}(a), where the solid lines represent constant values
of $r=f_n/f_p=r_{min}$, corresponding to ${\cal D}_{min}$ when the NLO
expression (\ref{eq:diff_rate_nlo}) is used instead. Indeed, values as
small as -2 are now possible\cite{isospin_violation_nlo2}. As already
pointed out, this is due to the fact that the cancellation in this
case is no longer between the WIMP couplings to protons and neutrons,
but between the latter ad the two--nucleon amplitude given in the last
of Eqs.(\ref{eq:nlo_amplitudes}). In the same figure the short--dashed
and long dashed straight lines represent the ``alignment'' conditions
given in Eqs.(\ref{eq:straight_line}) and
(\ref{eq:straight_line_nlo}), respectively.  As discussed in Section
\ref{sec:nlo}, close to these straight lines the WIMP expected rate
vanishes for all targets at the same time, so that, strictly speaking,
the compatibility ratio ${\cal D}$ cannot be minimized in the first
place. In practice close to those lines the $r_{min}$ parameter is
subject to large numerical oscillation when ${\cal D}$ is minimized.

As pointed out in Section \ref{sec:nlo}, however, the most relevant
quantity from the phenomenological point of view is the minimal
achievable value ${\cal D}_{min}$ of the ${\cal D}$ ratio, which in
the NLO case is not driven by the $r$ parameter, but instead by the
$t$ parameter defined by the scaling law recast as in
Eq.(\ref{eq:scaling_law_t}). Indeed, when the energy--dependent terms in the NLO
corrections of Eq.(\ref{eq:nlo_amplitudes}) are neglected, ${\cal D}$
is minimized to the same value ${\cal D}_{min}$ of the LO case, albeit
for a value of the $t$ parameter, $t_{min}\simeq -\bar{A}/Z$, which
corresponds to a different value of $r$ in each point of the parameter
space. This is shown if Fig.\ref{fig:r_fn_fp}(b), where the solid
lines represent constant values of $t_{min}$ when the full NLO
corrections (\ref{eq:nlo_amplitudes}) are included. From this figure
one can see that indeed, in large parts of the parameter space,
$t_{min}$ is very close to the constant value $\simeq -2.3$. The only
exception is close to the long-dashed straight line correspondent to
Eq.(\ref{eq:straight_line_nlo}), where the energy--independent part of
the amplitude cancels out so that the energy--dependent corrections
can no longer be neglected: it is this effect that leads to the
fluctuations in the values of $t_{min}$ found by the ${\cal D}$
minimization procedure. Notice that the energy dependence of the
scaling law is also expected to spoil the cancellation in the ${\cal
  D}$ minimization leading to higher values of ${\cal D}_{min}$ and in
this way playing against the possibility to make CDMS--$Si$ and
SuperCDMS mutually compatible.  This will be confirmed by our
numerical analysis.

As discussed in Section \ref{sec:isospin_violation}, the parameter
space close to the line of Eq.(\ref{eq:straight_line}) or
Eq.(\ref{eq:straight_line_nlo}) has also another important feature:
thanks to the factorization (\ref{eq:scaling_law_factorized}), for a
given target nucleus the degrading factor (\ref{eq:degrading_factor})
can become arbitrarily small, even in presence of many isotopes, so
that, if the expected WIMP rate is fixed to explain CDMS--Si, the
correspondent scale $M_*$ is driven to its smallest values. This is
confirmed by Fig.\ref{fig:lambda_theta_m_star}, where the solid lines
show constant values of $M_*$, calculated using
Eqs.(\ref{eq:lambda_tilde},\ref{eq:m_star_max}) with
$\rho_{\chi}$=$\rho_{local}$/2, with $\rho_{local}$=0.3 GeV/cm$^3$
(notice that we divide $\rho_{local}$ by two to be consistent with the
discussion of Section \ref{sec:relic_abundance}, where a scenario with
equal densities for the two states $\chi$ and $\chi^{\prime}$ is
outlined in which direct detection experiments are only sensitive to
$\chi$ down--scatters). As shown in Fig.\ref{fig:lambda_theta_m_star},
indeed the smallest values for $M_*$ are reached close to the straight
line (\ref{eq:straight_line_nlo}).  This suppression mechanism of
$M_*$ is expected to enhance both the annihilation cross section of
Eq.(\ref{eq:sigmav}) and the LHC signals: we wish now to analyze this
in detail combining the discussion of the direct detection signal
(Section \ref{sec:lambda_from_data}) with the relic abundance
calculation (Section \ref{sec:relic_abundance}) and signals at the LHC
(Section \ref{sec:LHC}).

The Lagrangian of Eq. (\ref{eq:eff_lagrangian_full}) depends on the 6
couplings $\tilde{\lambda}_q$ ($q=u,d,s,c,b,t$) and, as discussed in
Section \ref{sec:model}, the phenomenology is expected to depend on
the five ratios $\tilde{\lambda}_q/\tilde{\lambda}_u$
($q=d,s,c,b,t$). At variance with the other observables, however,
direct detection is sensitive to scales much lower than that of heavy
quarks, so that the latter are integrated out and only enter in the
calculation of the expected rate through the combination
$\lambda_{\theta}$=2/27$\sum_{Q=c,b,t}\tilde{\lambda}_Q$. This implies
that in each point of the plane
$\bar{\lambda}_{\theta}$--$\bar{\lambda}_s$ only the sum of
heavy-quark couplings is fixed by direct detection, while, in order to
calculate other observables, all the couplings $\tilde{\lambda}_Q$ are
needed. Notice, however, that our choice of the IDM parameters
corresponds to $m_{\chi}$,$m_{\chi^{\prime}}$ $\lsim m_b$. This means
that in the annihilation cross section $<\sigma v>$ only the
annihilation channels $u\bar{d}$, $d\bar{d}$, $s\bar{s}$ and
$c\bar{c}$ are kinematically accessible. As already mentioned in
Section \ref{sec:relic_abundance}, in the isospin--conserving case the
Lagrangian of Eq.(\ref{eq:eff_lagrangian_full}) leads to a $p$--wave,
velocity--suppressed $<\sigma v>$ that drives the thermal relic
abundance above the observational constraints. In the following we
wish to explore the possibility that the IVDM mechanism may instead
allow to find values of the thermal relic abundance compatible to
observation, so that we are interested in maximizing $<\sigma v>$. In
light of this, in the following we will fix
$\tilde{\lambda}_b$=$\tilde{\lambda}_t$=0, so that in each point of
our parameter space $\tilde{\lambda}_c$=27/2 $\lambda_{\theta}$.

The result of a combined analysis of the relic abundance and of the
minimum compatibility ratio ${\cal D}_{min}$ between CDMS--Si and
SuperCDMS is shown in Figs. \ref{fig:allowed_region},
\ref{fig:eta_vmin} and \ref{fig:degrading_factor}.  Figure
\ref{fig:allowed_region} shows the
$\bar{\lambda}_{\theta}$--$\bar{\lambda}_{s}$ parameter space, where
again the short--dashed and long dashed straight lines represent
Eqs.(\ref{eq:straight_line}) and (\ref{eq:straight_line_nlo}),
respectively. In this Figure the shaded regions represent the
parameter space where CDMS--Si and SuperCDMS are mutually compatible
while at the same time the metastable state $\chi$ can be a thermal
relic (i.e. $\Omega_{\chi}h^2\le$0.12 using Eq. (\ref{eq:omega})) and
are bounded by the solid blue lines which correspond to the two
conditions $\Omega_{\chi}h^2$=0.12 and ${\cal D}_{min}$=1.  Here in
the evaluation of ${\cal D}$ the expected WIMP signal has been
calculated using Eq.(\ref{eq:diff_rate_nlo}), i.e. including the NLO
corrections of
Ref.\cite{isospin_violation_nlo,isospin_violation_nlo2}.  The
existence of such a region in the parameter space, close to the values
fixed by Eq.(\ref{eq:straight_line_nlo}) but not overlapping them, is
the main result of our analysis. In the same figure the dotted curve
represents $\Omega_{\chi}h^2$=0.12 calculated using the LO scaling law
for the expected rate (see Eq.(\ref{eq:diff_rate})). This curve is
only marginally modified compared to the LO case and confirms what we
already pointed out: with very few exceptions the phenomenology is
only slightly modified by NLO corrections in spite of the fact that
$r_{min}$ can be sizeably changed. Finally, the inner solid (red) line
corresponds to $\tau=1/\Gamma$=4$\times$10$^{26}$ seconds, as given by
Eq.(\ref{eq:decay}): indeed for such low values of the $|\delta|$
parameter \cite{decay} the lifetime of the metastable state $\chi$ is
much larger than the age of the Universe, $\tau_U\simeq 4.35 \times
10^{17}$ seconds. In Figure \ref{fig:allowed_region} the (red) circle
indicates the representative choice of $\bar{\lambda}_{\theta}$,
$\bar{\lambda}_{s}$ for which measurements and bounds for the function
$\tilde{\eta}$ defined in Eq.(\ref{eq:eta_tilde}) with
$\sigma_0=\sigma_p$ are shown in detail in
Fig.\ref{fig:eta_vmin}(a). This choice corresponds to
$\Omega_{\chi}h^2$=0.12 and ${\cal D}_{min}$=0.7. On the other hand,
the (red) square indicates the choice of $\bar{\lambda}_{\theta}$,
$\bar{\lambda}_{s}$ for which $\tilde{\eta}$ is discussed in
Fig.\ref{fig:eta_vmin}(b), while ${\cal D}$ is plotted as a function
of $\bar{\lambda}_d$ in Fig. \ref{fig:degrading_factor}(a). In this
case ${\cal D}_{min}$=1, i.e. this configuration is at the verge of
incompatibility between the CDMS--Si result and the SuperCDMS
constraint: consistently with the condition ${\cal D}_{min}$=1, in
Figure \ref{fig:eta_vmin}(b) the constraint from SuperCDMS ``touches''
the upper range of the CDMS--$Si$ excess.  Finally, in the same
figure, the (red) triangle exemplifies a configuration very close to
Eq.(\ref{eq:straight_line_nlo}) where the NLO energy--dependent
corrections of Eq. (\ref{eq:diff_rate_nlo}) spoil the maximal
achievable cancellation in WIMP--$Ge$ scattering. For illustrative
purposes the corresponding compatibility ratio ${\cal D}$ is plotted
in Fig.\ref{fig:degrading_factor}(b).

As expected, getting close to the straight line of
Eq. (\ref{eq:straight_line_nlo}) leads to two opposite effects: on the
one hand $M_*$ is suppressed, driving the thermal relic density
$\Omega_{\chi} h^2$ down to values compatible to observation; on the
other, it suppresses the energy--independent part of the scattering
amplitude, enhancing the r\"{o}le of energy--dependent NLO corrections
and spoiling the cancellation in the compatibility factor, so that
${\cal D}_{min}$ can become larger then unity. The extent of this
second effect is shown quantitatively in
Figs. \ref{fig:degrading_factor}(a,b), where the compatibility ratio
${\cal D}$ defined in Eq.(\ref{eq:degrading_factor_eta_i}) is plotted
as a function of $\bar{\lambda}_d$ in the two representative cases
$\bar{\lambda}_{\theta}$=-0.1, $\bar{\lambda}_s$=1 and
$\bar{\lambda}_{\theta}$=-0.1, $\bar{\lambda}_s$=1.4,
respectively. These two configurations are the benchmark points
indicated with a (red) square and triangle, respectively, in
Fig.\ref{fig:allowed_region}. In both plots of
Fig.\ref{fig:degrading_factor} the solid (blue) line represents the
calculation including the NLO corrections of
Eq. (\ref{eq:diff_rate_nlo}), while the thin dotted (black) line shows
the same quantity when the approximate expression of
Eq.(\ref{eq:diff_rate_nlo_approx}) for the NLO corrections is used,
which neglects the terms with explicit energy dependence. Both plots
show in a clear way how these latter terms are instrumental in driving
${\cal D}_{min}$ above unity.

We conclude our discussion showing in Fig.\ref{fig:lhc} some
predictions for LHC signals. In particular, the thin solid lines
represent constant values of the expected number of monojet+missing
energy events for the integrated luminosity ${\cal L}$=19.5 $fb^{-1}$,
and ranges from 100 to 1000 events. As a reference, for the same
integrated luminosity and kinematic cuts Ref. \cite{pt_jet} claims an
upper bound of about 400 events for the same quantity. On the other
hand, in the same figure the thick solid lines represent constant
values of the cross section for hadronically--decaying mono $W/Z$
events, ranging from 100 to 500 $fb$. Since the corresponding 95\%
C.L. upper bound on the same quantity is 4.4 $fb$ \cite{atlas_wz},
this latter result appears to be in strong tension with
observation. Notice, however, that the validity of the Effective Field
Theory approach is questionable when the momentum exchanged in the
propagator driving the process is of the same order of the suppression
scale $M_*$ or larger \cite{eft_validity_lhc}: indeed, this appears to be
the case from the values of $M_*$ shown in
Fig.\ref{fig:lambda_theta_m_star}.

\section{Conclusions}
\label{sec:conclusions}

In the present paper we have explored a specific scenario of light
Inelastic Dark Matter (IDM) with $m_{\chi}\lsim4$ GeV and $\delta<0$
(exothermic DM) where the couplings violate isospin symmetry (IVDM)
leading to a suppression of the WIMP cross section off Germanium
targets.  This combination of IDM and IVDM parameters, which allows to
find compatibility between an explanation of the CDMS--Si excess in
terms of WIMP scatterings and constraints from LUX, XENON100 and
SuperCDMS, has been discussed by several authors\cite{ge_phobic,noi}.
We have extended the existing analyzes in different directions in the
case of an Effective Field Theory model for a Dirac
  IDM particle with a scalar coupling to quarks:
\begin{itemize}
\item we have fully incorporated the halo--independent approach by
  introducing an appropriately defined compatibility ratio (see
  Eq.(\ref{eq:degrading_factor_eta_i});
\item we have explored the isospin--violating coupling constant
  parameter space to discuss the maximal achievable degrading factors
  within the IVDM scenario as well as the minimal values of the
  suppression scale $M_*$ required to explain the three CDMS--Si
  events in terms of WIMP scatterings;
\item we have discussed the effect on such an analysis of the
  inclusion of the NLO corrections recently discussed in
  \cite{isospin_violation_nlo,isospin_violation_nlo2};
\item we have included a discussion on the thermal relic density of
  the metastable state $\chi$, showing in which circumstances it can
  be compatible to observation;
\item we have also discussed accelerator bounds by showing that Large
  Hadron Collider (LHC) constraints from monojet and
  hadronically-decaying mono-W/Z searches can be severe for this
  scenario, although the application of EFT at the LHC is questionable
 given the ranges of the $M_*$ suppression scale parameter required
  by our analysis.
\end{itemize}

The main result of our analysis is that a region in the parameter
space exists (close to the straight line of
Eq. (\ref{eq:straight_line}) in the LO case or
Eq.(\ref{eq:straight_line_nlo}) in the NLO case) where WIMP
scatterings can explain the CDMS--$Si$ excess in compliance with other
experimental constraints, while at the same time the metastable state
$\chi$ can be a thermal relic.  This is at variance with what usually
happens for a fermionic DM particle with a scalar coupling to quarks
in the isospin--conserving case \cite{cdms_si_eff}.  In this scenario
the metastable state $\chi$ and the lowest--mass particle
$\chi^{\prime}$ have approximately the same density in the present
Universe and in our Galaxy, but direct detection experiments are only
sensitive to the down--scatters of $\chi$ to $\chi^{\prime}$. In
particular, we have shown that for this choice of parameters,
indicated with the shaded area in Fig.\ref{fig:allowed_region}, two
opposite effects are at work: on the one hand the effective scale
$M_*$ is suppressed, driving the thermal relic density $\Omega_{\chi}
h^2$ down to values compatible to observation, because the scaling law
acquires a factorization in terms of the couplings (see
Eq.(\ref{eq:scaling_law_factorized})) that allows the scattering
amplitude to become arbitrarily small also in presence of many
isotopes; on the other hand, when the parameters get too close to
Eq.(\ref{eq:straight_line_nlo}) energy--dependent NLO corrections can
spoil the cancellation in the compatibility factor, leading eventually
to tension between CDMS--Si and SuperCDMS (for a particular example
the extent of the latter effect is explained in detail in
Fig.\ref{fig:degrading_factor}).

We remind that NLO corrections to WIMP--nucleus scattering are
affected by sizable uncertainties, since some of them are only known
for nuclei with closed shells and a rough extrapolation is needed to
apply the formalism to nuclei used in real--life experiments,
including $Si$ and $Ge$ \cite{isospin_violation_nlo,
  isospin_violation_nlo2}. Nevertheless our conclusions that NLO
corrections are only relevant for the phenomenology in the couplings
parameter space close to
Eqs.(\ref{eq:straight_line},\ref{eq:straight_line_nlo}) is
qualitatively robust. In particular, we found that, with that notable
exception, the IVDM phenomenology is only slightly modified by NLO
corrections, in spite of the fact that the ratio between WIMP
couplings to neutrons and protons, $r=f_n/f_p$, which is required to
minimize the degrading factor between Silicon and Germanium can be
sizeably changed compared to the LO case.

\acknowledgments
This work was supported by the National Research
Foundation of Korea(NRF) grant funded by the Korea government(MOE)
(No. 2011-0024836).

\end{document}